%% file: main.tex
\documentclass[a4]{article}
\usepackage{arxiv}
\usepackage[utf8]{inputenc} 
\usepackage[T1]{fontenc}    
\usepackage{booktabs}       
\usepackage{amsfonts}       
\usepackage{nicefrac}       
\usepackage{microtype}      
\usepackage{textcomp}       
\usepackage{mathtools}

\usepackage{tikz}
\usetikzlibrary{positioning}

\usepackage{times}
\usepackage{graphicx}
\usepackage{subcaption}                 
\usepackage{amssymb}
\usepackage{url}
\usepackage{hyperref}
\usepackage{geometry}
\usepackage{amsmath}
\usepackage{physics}
\usepackage{verbatim}                   
\usepackage[ruled,vlined]{algorithm2e}  
\usepackage{fancyhdr}
\usepackage{cleveref}   

\usepackage{listings}

\usepackage{color}
\definecolor{gray}{rgb}{0.4,0.4,0.4}
\definecolor{darkblue}{rgb}{0.0,0.0,0.6}
\definecolor{cyan}{rgb}{0.0,0.6,0.6}
\definecolor{backcolour}{rgb}{0.95,0.95,0.92}

\lstdefinelanguage{XML}
{
  morestring=[b]",
  morestring=[s]{>}{<},
  morecomment=[s]{<?}{?>},
  stringstyle=\color{black},
  identifierstyle=\color{darkblue},
  keywordstyle=\color{cyan},
  captionpos=b,
  backgroundcolor=\color{backcolour},
  numberstyle=\tiny\color{gray},
  numbers=left,                    
  numbersep=5pt,
  morekeywords={xmlns,version,type}
}
\crefname{lstlisting}{listing}{listings}
\Crefname{lstlisting}{Listing}{Listings}

\graphicspath{{figures/}}

\fancypagestyle{firststyle}
{
   \fancyhf{}
   \chead{}
}

\title{L-QLES: Sparse Laplacian generator for evaluating Quantum Linear Equation Solvers}
\author{Leigh Lapworth \\
\\
Rolls-Royce plc \\
Derby, UK \\
\today
\\
\\
leigh.lapworth@rolls-royce.com  \\
}

\begin{document}

\maketitle

\begin{abstract}
L-QLES is an open source python code for generating 1D, 2D and 3D Laplacian operators 
and associated Poisson equations and their classical solutions. 
Its goal is to provide quantum algorithm developers with a flexible test case 
framework where features of industrial applications can be incorporated without
the need for end-user domain knowledge or reliance on inflexible one-off
industry supplied matrix sets.
A sample set of 1, 2, and 3 dimensional Laplacians are suggested and 
used to compare the performance of the Prepare-Select and FABLE block 
encoding techniques.
Results show that large matrices are not needed to investigate industrial
characteristics. A matrix with a condition number of 17,000 can be encoded 
using 13 qubits.
L-QLES has also been produced to enable algorithm developers to investigate
and optimise both the classical and quantum aspects of the inevitable
hybrid nature of quantum linear equation solvers. 
Prepare-Select encoding that takes over an hour of classical preprocessing time
to decompose a 4,096x4,096 matrix into Pauli strings can be can investigated 
using L-QLES matrices. 
Similarly, row-column query oracles that have success probabilities $\le 10^{-7}$
for the same matrix can be investigated.
\end{abstract}

\input{include/intro}
\input{include/laplacians}
\input{include/l-qles}

\input{include/testcases}
\input{include/results}

\input{include/close}

%
\newpage
\bibliographystyle{ieeetr}
\bibliography{references}

%
\newpage
\appendix
\input{include/appendix-ksum}
\input{include/appendix-cluster}

\input{include/appendix-lqles}

\end{document}

%% file: include/intro.tex
%
\section{Introduction}
\label{sec-intro}

A recent review of opportunities for quantum advantage \cite{hoefler2023disentangling},
found that {\it a wide range of often-cited applications 
is unlikely to result in a practical quantum advantage without significant algorithmic improvements.}
Does this presage a global wave of disillusionment amongst expectant beneficiaries?
Or, is it a call to arms for algorithm developers? It maybe both. 
But without the latter, the former becomes a certainty.

Computational Fluid Dynamics (CFD) is one of the areas where a quantum advantage is
far from guaranteed. Indeed, the wider discipline of sparse matrix algebra falls into the
same category.
The development of industrial CFD codes benefitted greatly from benchmark test cases such 
as NASA rotor 37 \cite{denton1997lessons, reid1978design}
for turbomachinery flows and the sequence of AIAA drag prediction test cases and workshops
\cite{levy2002summary, tinoco2018summary} for external aerodynamics. 
Each case has high quality experimental data for developers to verify the accuracy of their codes. 

One key feature of the benchmark test cases is that they have driven complexity: 
more accurate answers need better meshes, better discretisation schemes, 
better turbulence models etc.
Quantum Linear Equation Solvers (QLES) do not need to repeat that journey, but they
do need to embrace the complexity of industrial usage.
Therein lies a practical difficulty. 
Quantum algorithm developers do not, generally, have the domain expertise to generate 
relevant matrices from high level test case descriptions.
Industry can provide sample matrices \cite{lapworth2022hybrid} but these often present
developers with a single bound to traverse rather than a series of incremental steps.

This work provides algorithm developers with a test case framework, L-QLES, that allows
them to build their own sequence of QLES test matrices which have some of the characteristics
of industrial applications.
L-QLES seeks to bridge the gap between industrial complexity and what is feasible on
near and medium term devices. It is a simple Poisson equation solver with a focus on
creating Laplacians of varying types and degree of difficulty.
It can incorporate typical mesh distributions, different types of boundary condition
and matrix structure. Sample test parameters are provided.
These include matrices whose condition numbers range from $O(10)$ to $O(10^5)$
and a permutation operator whose application results in the number of Pauli strings 
in the decomposition of the Laplacian increasing by a factor of 30.

This preprint is organised as follows: the next section gives a brief background to 
Laplacian operators. The high level capabilities of L-QLES are then described.
Sample 1, 2 and 3 dimensional test cases are then introduced each with details of
their test conditions.
To illustrate the use of L-QLES the test Laplacians are used to compare the performance
to two different matrix encoding approaches \cite{childs2017quantum} and 
\cite{camps2022fable}.

See \Cref{sec-data} for the availability of L-QLES and the input files for the 
sample test cases.
 

%% file: include/laplacians.tex
%
\section{Laplacian operators}
\label{sec-laplacian}

The inhomogeneous Laplace equation (i.e. Poisson equation) is:

\begin{equation}
    \nabla^{2} \phi = f(\Vec{x})
    \label{eqn-poisson1}
\end{equation}

where $\nabla$ is the differential operator over a given $d$-dimensional 
volume $\Omega \subset \mathbb{R}^d$, $f$ is a scalar force that depends
on the position $\Vec{x}$ in $\Omega$, and $\phi$ is the solution state.
\Cref{eqn-poisson1} is supplemented with boundary conditions on
$\partial \Omega$. 

Here, \Cref{eqn-poisson1} is discretised over 1, 2 and
3 space dimensions using a lattice based mesh. 
The resulting discrete equation is written as:
\begin{equation}
    L \ket{\phi} = \ket{b}
\end{equation}

where $L$ is an $N \times N$ matrix and $\ket{\phi}$ are $\ket{b}$ are
N dimensional state vectors representing the discrete solution and the
inhomogeneous terms from the scalar force and the boundary conditions.
$N$ is the total number of points on the lattice. 
For example, a 16x16 lattice in 2-dimensions has $N=256$.

The use of the Poisson equation is common in quantum algorithms research.
This is not unexpected as its elliptic nature makes it more compute intensive to 
solve on classical computers than parabolic convection equations.
Typically, 1-dimensional analyses use a simple Laplacian with row-wise entries
of the form (-1, 2, -1)
\cite{liu2021variational, sato2021variational, ali2023performance}.
In 2-dimensions, the regular form (-1, -1, 4, -1, -1) is often adopted
\cite{daribayev2023implementation}.
A more general 2D formulation \cite{sunderhauf2024block} has been used to
demonstrate low depth circuit encoding by exploiting repeated elements in
the matrix.
A 2D Laplacian has also been formed from the Kronecker sum of 1D Laplacians
\cite{camps2022fable}:

\begin{equation}
    L = L_{yy} \oplus L_{xx} = I \otimes L_{xx} + L_{yy} \otimes I
\end{equation}

For 2D Cartesian lattice meshes, as will be used here,
this produces the correct rows of the Laplacian for interior points
of the lattice, but it only produces the correct boundary rows when repeat
boundary conditions are used in both directions, see \Cref{sec-ksum-vs-disc}.
As \Cref{subsec-testcases-2d-RRRR} shows, there are cases of interest
where the Kronecker sum approach produces the correct Laplacian.

Although industrial sparse matrix applications are predominantly non-linear
and the performance of QLES can only be fully assessed within the context of an 
outer non-linear iterative scheme \cite{lapworth2022implicit},
there are important features that can be investigated with a Poisson solver.
Algorithms such as HHL \cite{harrow2009quantum} and QSVT \cite{gilyen2018quantum}
scale with the matrix condition number.
In the sample test cases in \Cref{sec-testcases}, matrices with condition
numbers up to 500,000 are created.
A 64x64 matrix with a condition number of 17,000 that can be inverted using QSVT 
with 14 qubits is already a significant challenge due to the number of phase 
factors needed \cite{dong2021efficient}.

%% file: include/l-qles.tex
%
\section{L-QLES}
\label{sec-l-qles}

L-QLES is an open source Python code for generating 1D, 2D and 3D Laplacian
operators and associated Poisson equations and their classical solutions. 
The Laplacians are created using a finite volume discretisation
\cite{darwish2016finite} on Cartesian lattice meshes. 
The key feature of L-QLES is the ability to {\it tune} 
the mesh and, hence, the Laplacian to include the following features of 
industrial applications:

\begin{itemize}
    \item Non-uniform mesh distributions,
    \item Multiple boundary condition types,
    \item Arbitrary mesh indexing.
\end{itemize}

The mesh and boundary conditions are set via a single XML input file.
\Cref{lst-input1} shows the input file for a 1D mesh. 
2D and 3D meshes are created by changing {\it dimension} in line 3 and
adding the equivalent {\it mesh} sections for the "y" and "z" directions.

\begin{lstlisting}[language=XML, label=lst-input1, caption=XML input file for 1D Laplacian]
<?xml version="1.0" encoding="UTF-8"?>
<laplace>
  <case name="l1d_16_dd" dimension="1" force="1.0"></case>
  <mesh direction="x">
    <length>1.0</length>
    <ntotal>16</ntotal>
    <nclust>6</nclust>
    <cltype>2</cltype
    <cratio>1.2</cratio> 
    <btype>D, D</btype>
    <bvalue>0.0, 0.0</bvalue>
    <degfix>8</degfix>
  </mesh>
</laplace> 
\end{lstlisting}

See \Cref{sec-run-lqles} for the command line arguments to run L-QLES.

%
\subsection{Non-uniform meshes}
\label{subsec-nonu-mesh}
The most common reason to use non-uniform meshes is to resolve regions
of high gradients such as viscous boundary layers in flows along walls.
These are most pronounced in turbulent flows which require the Navier-Stokes
equations to be modelled. L-QLES provides representative non-uniform meshes
without the non-linearities of the Navier-Stokes equations.

Second order finite volume codes typically require a smooth variation in
spacing from very small mesh cells next to each wall and a larger mesh spacing 
away from the walls. 
The algorithm for generating the mesh coordinates from the input
parameters in \Cref{lst-input1} is described in \Cref{sec-clustering}.

%
\subsection{Boundary conditions}
\label{subsec-bcs}

The boundary condition types support by L-QLES are listed below.
Each is specified by its capitalised first letter and a boundary condition
must be specified for each end of the domain, as in line 10 of
\Cref{lst-input1}. As on line 11, a boundary value must also be
specified for each end of the domain.

\begin{description}
\item [Dirichlet ('D')] The value of the solution, $\phi$, at the boundary point is
      specified. 
      If $i$ is the row index corresponding to the boundary point, the off-diagonal
      entries on the $i^{th}$ row are set to zero and the boundary value is modified to give the boundary equation: $a_{ii}\phi_i=a_{ii}b_i$.
      
\item [Neumann ('N')] The first derivative of the solution at the 
      boundary point is specified. 
      If j is the index of the immediate neighbour to the boundary, this
      condition is crudely implemented by setting the off-diagonal
      entries on the $i^{th}$ row  to zero except for setting
      $a_{ij} = -a_{ii}$. This gives the boundary equation: 
      $a_{ii}(\phi_i-\phi_j)=A_i b_i$, where $b_i$ is the required boundary 
      gradient. $A_i$ is the area of the boundary face and 
      $a_{ii} = A_i/dx_i$ for an $x$ boundary.
      Although crude, this is correct for the Cartesian lattice meshes
      used by L-QLES.

\item [Repeat ('R')] The domain is cyclic with a repeating solution.
      In 1-dimensional, when discretising the Laplacian at node $\phi_0$
      there is a {\it fictitious} volume and node $\phi_{-1}$ such that
      $\phi_{-1} = \phi_{nx-1}$. Similarly, there is a {\it fictitious}
      $\phi_{nx} = \phi_{0}$.
      Note that this form of repeat boundary is appropriate for cell-centred
      discretisation schemes and is the form used by \cite{camps2022fable}.
      The alternative form that sets the boundary values equal is not used here.
      
\item [Symmetry ('S')] This is same as the Neumann boundary condition
      with the boundary gradient enforced to be zero. This can also be achieved
      by setting the boundary gradient to zero in the Neumann condition.
\end{description}

%
\subsection{Arbitrary mesh indexing}
\label{subsec-reorder}
In classical sparse matrix algebra, computational performance can be significantly
enhanced using techniques such as bandwidth reduction \cite{cuthill1969reducing} 
and cache optimised reordering via space-filling curves \cite{aftosmis2004applications}.
Whilst these techniques may not be relevant to QLES, Laplacians relevant to
industrial applications do not have coordinate based mesh indexing.
To investigate irregular mesh indexing, L-QLES provides a simple {\it shell} based 
reordering.

In 3D, the shell ordering starts at index (0,0,0) and orders the indices by
alternately taking one step in each mesh direction until a boundary is reached.
This creates a mapping between the two orderings which is used to define two
permutation matrices $P$ and $Q$ which, respectively reorder the 
rows and columns of the matrix. Noting that $P^{-1}=P$ and $Q^{-1}=Q$,
the discrete Poisson equation $L\ket{\phi}=\ket{b}$ becomes:

\begin{equation}
    PLQ.Q\ket{\phi} = P\ket{b}
    \label{eqn-reorder}
\end{equation}

\Cref{fig-l3d_4x8x8} shows the original and reordered sparsity patterns for
a 3-dimensions 4x8x8 mesh. The reordering is activated by a command line 
argument. See \Cref{sec-run-lqles}

%
\subsection{Inhomogeneous RHS term}
\label{subsec-force}
Since the focus of L-QLES is on the structure of the Laplacian, the
inhomogeneous force term on the RHS of the Poisson equation is set as a 
constant via the value of {\it force} on line 3 of \Cref{lst-input1}.

%
\subsection{Illustration}
\label{subsec-illust}

\Cref{tab-illust1d} gives an illustration of how the mesh parameters in L-QLES can
be used to vary the condition number, $\kappa$, of the resulting Laplacian.
There is no direct way to enforce a desired condition number, but a handful of trial 
and error runs should suffice.
Note that the classical time to compute the condition number scales with $O(N^3)$ where
$N$ is the dimension of the matrix. Here, the condition number is the ratio of the
largest and smallest eigenvalues.

\begin{table}[h]
  \centering
  \begin{tabular}{c c c c c c c }
    \toprule
    $n_t$ & $n_c$ & $r$  &  $\kappa$   \\
    \midrule
     8    &   3   & 1.3  & 16      \\
     16   &   6   & 1.3  & 107      \\
     16   &   6   & 2    & 376      \\
     32   &   7   & 1.3  & 779      \\
     32   &   12  & 1.3  & 1224      \\
     32   &   12  & 1.6  & 6,636     \\
     64   &   18  & 1.25 & 17,006    \\
     128  &   43  & 1.1  & 70,635    \\
    \bottomrule \\
  \end{tabular}
  \caption{1D Laplacian with Dirichlet boundary conditions, variable in condition
           number with different L-QLES settings.}
  \label{tab-illust1d}
\end{table}

%% file: include/testcases.tex
%
\section{Test cases}
\label{sec-testcases}
Whilst the intention of L-QLES is to enable users to configure their
own test cases, reference cases are useful for cross comparisons.
In this section, test cases are recommended that have some elements
of industrial usage.
The input files for a range of test cases, including all those used here, are
available on-line.
\footnote{\href{https://github.com/rolls-royce/qc-cfd/tree/main/L-QLES/input_files}{https://github.com/rolls-royce/qc-cfd/tree/main/L-QLES/input\_files}}

%
\subsection{1D test cases}
\label{subsec-testcases-1d}

%
\subsubsection{Poiseuille flow}
\label{subsec-testcases-1d-pois}

Poiseuille flow is the fully developed flow in a semi-infinite 2D wall 
bounded channel and is governed by the equation:
\begin{equation}
  \frac{\partial}{\partial x} \mu \frac{\partial u}{\partial x} = \frac{\partial p}{\partial y}
  \label{tests_poiseuille01}
\end{equation}

Where both the laminar viscosity, $\mu$, and the pressure gradient, $\frac{\partial p}{\partial y}$ in
the flow direction, are both constants:

\begin{equation}
  \frac{\partial^2 u}{\partial x^2} = k = \frac{1}{\mu}\frac{\partial p}{\partial y}
  \label{tests_poiseuille02}
\end{equation}

The wall boundary conditions are typically $u(0)=u(L)=0$ and the analytic
solution is:
\begin{equation}
  u = \frac{k}{2}x^2 - \frac{kL}{2} x 
  \label{tests_poiseuille03}
\end{equation}

%
\subsubsection{Steady heat equation}
\label{subsec-testcases-1d-heat}
The 1-dimensional heat equation is:
\begin{equation}
  \frac{\partial u}{\partial t} =
  -\frac{1}{c_p}
  \frac{\partial}{\partial x} \left(-k \frac{\partial u}{\partial x} \right)
  \label{tests_heat01}
\end{equation}
For steady flow with constant thermal conductivity, $k$ and specific heat
capacity, $c_p$ this reduces to:
\begin{equation}
  \frac{\partial^2 u}{\partial x^2} =0
  \label{tests_heat02}
\end{equation}

The boundary conditions are usually set to non-equal non-zero values.

%
\subsubsection{Test conditions}
\label{subsubsec_test_conds_1d}

Suggested test conditions are listed in \Cref{tab-tests1d}. 
These are ordered to give an increasing condition number ranging from
$O(10)$ to $O(10^4)$. 
For a given total number of points, increasing the number of points in
the clustered region and/or the expansion ratio increases the
condition number.

\begin{table}[h]
  \centering
  \begin{tabular}{c c c c c c c }
    \toprule
    \multicolumn{2}{c}{} & \multicolumn{2}{c}{Dirichlet} & \multicolumn{2}{c}{Repeat} \\
    $n_t$ & $n_c$ & $r$ & $\kappa$ & $r$ & $\kappa$  \\
    \midrule
     8    &   3   & 1.3  & 16     & 1.0 & 25 \\
     16   &   6   & 1.3  & 107    & 1.0 & 103 \\
     32   &   12  & 1.2  & 735    & 1.0 & 414 \\
     64   &   22  & 1.2  & 13,054 & 1.0 & 1,659 \\
     128  &   43  & 1.1  & 70,636 & 1.0 & 6,639 \\
    \bottomrule \\
  \end{tabular}
  \caption{ Suggested test conditions for 1D Dirichlet and Repeat Laplacians.}
  \label{tab-tests1d}
\end{table}

\begin{figure}[ht]
  \centering
  \captionsetup{justification=centering}
    \begin{subfigure}{0.4\textwidth}
    \includegraphics[clip, trim=0.0cm 0cm 0.0cm 1.35cm, width=\textwidth]{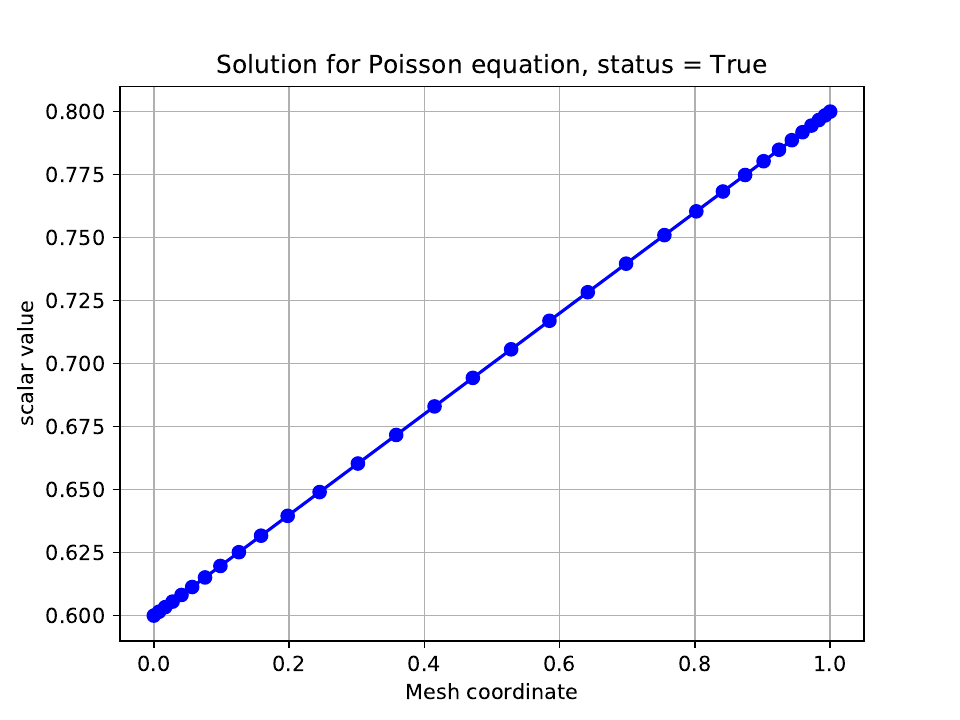}
    \caption{Heat equation with $u(0)=0.6$ and $u(L)=0.8$.}
    \label{fig-heat-sol}
  \end{subfigure}
  \captionsetup{justification=centering}
    \begin{subfigure}{0.4\textwidth}
    \includegraphics[clip, trim=0.0cm 0cm 0.0cm 1.35cm, width=\textwidth]{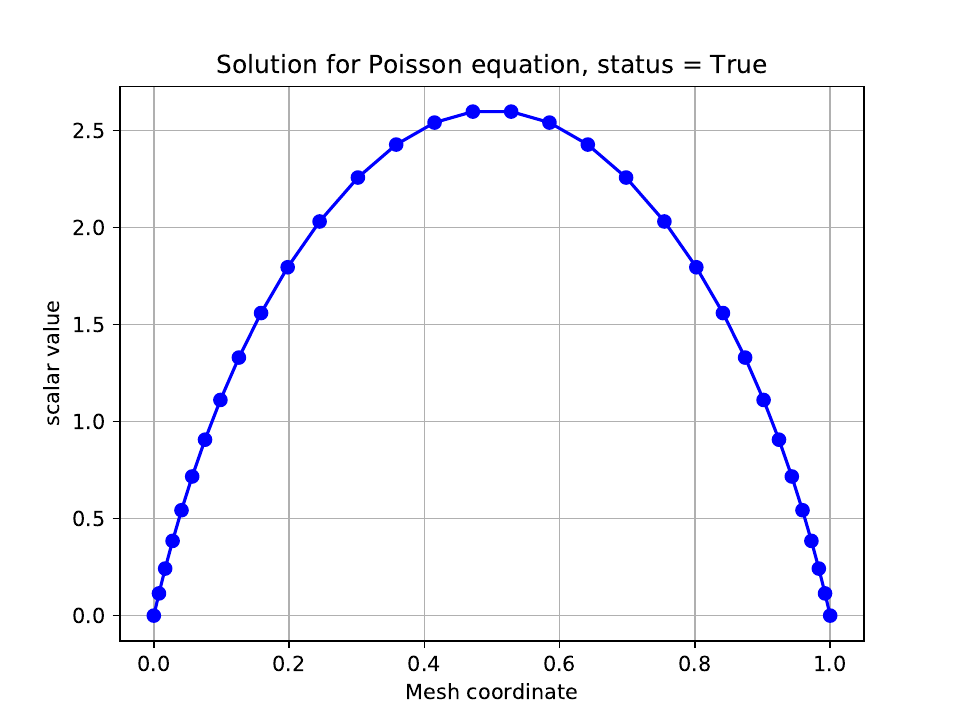}
    \caption{Poiseuille flow with $k=1$.}
    \label{fig-Poiseuille-sol}
  \end{subfigure}
  \caption{Classical solutions for Heat equation and Poiseuille test cases with
           $n_t=32$, $n_c=12$ and $r=1.2$.}
  \label{fig-solutions}
\end{figure}

The classical solutions for the Poiseuille flow and heat equations 
using the $n_t=64$ mesh are shown in \Cref{fig-solutions}.

%
\subsection{2D test cases}
\label{subsec-testcases-2d}
The 2D test conditions are chosen to exercise the range of
boundary conditions.

%
\subsubsection{Double repeating}
\label{subsec-testcases-2d-RRRR}

\begin{figure}[ht]
  \centering
  \captionsetup{justification=centering}
  \includegraphics[clip, trim=1.0cm 1cm 1.0cm 0.5cm, width=0.72\textwidth]{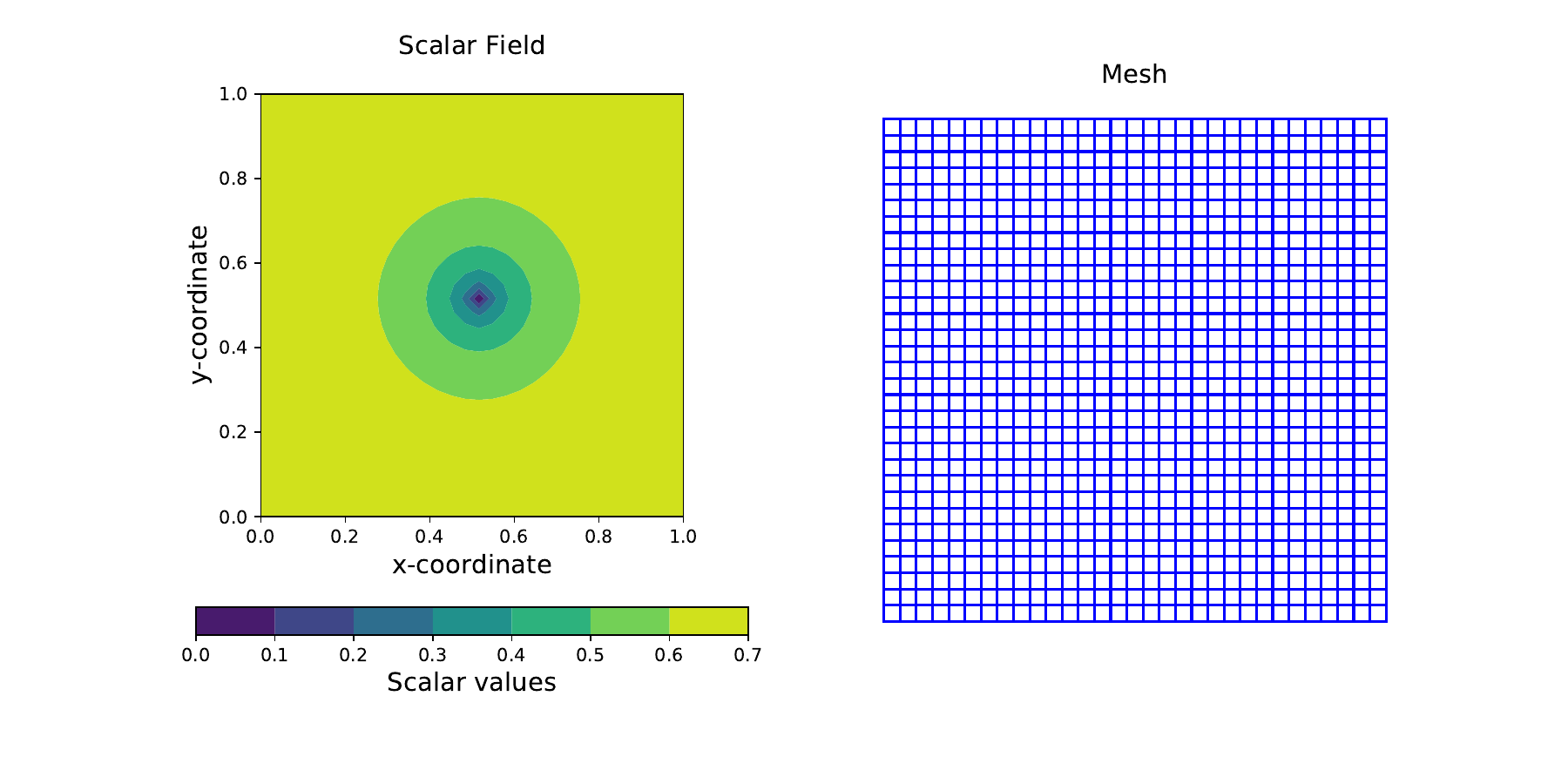}
  \caption{Solution and mesh for the solution of a 2D Laplacian with double
           repeating boundary conditions on 32x32 mesh.}
  \label{fig-l2d_32x32_rrrr}
\end{figure}

Double repeating boundary conditions in the $x$ and $y$ directions
are representative of modelling identical Taylor Green vortices 
\cite{taylor1937mechanism} in an unbounded 2D domain.
The actual modelling of the vortices requires the full 
Navier-Stokes equations. 
However, the Laplacian operator is indicative of the Poisson 
equation used to solve for the pressure in an incompressible 
CFD solver. In these cases the mesh is uniform as shown 
in \Cref{fig-l2d_32x32_rrrr}.

Without any modification, the double repeating Laplacian is
degenerate and cannot be solved classically. 
If the solution is part of a pressure correction solver, then it
is only gradients of the solution that are important and the Laplacian
can be modified at a single point (i.e. row) to have a Dircichlet 
condition that removes the degeneracy.
The choice of point is an arbitrary user input via the setting
of {\it degfix} on line 12 of \Cref{lst-input1}.
In \Cref{fig-l2d_32x32_rrrr}, the fixed point is set at the 
coordinates (16,16).

For quantum solvers, the normalisation of the state vector also
removes the degeneracy and L-QLES has the option to
generate Laplacians with and without the degeneracy.
As shown in \Cref{sec-ksum-vs-disc}, the degenerate Laplacian is the
same when generated by Kronecker sums or direct discretisation.
This is the only case for which this is true and provides a 
useful case for comparing with previous work.

%
\subsubsection{Double Neumann}
\label{subsec-testcases-2d-NNNN}

\begin{figure}[ht]
  \centering
  \captionsetup{justification=centering}
  \includegraphics[clip, trim=1.0cm 1cm 1.0cm 0.4cm, width=0.66\textwidth]{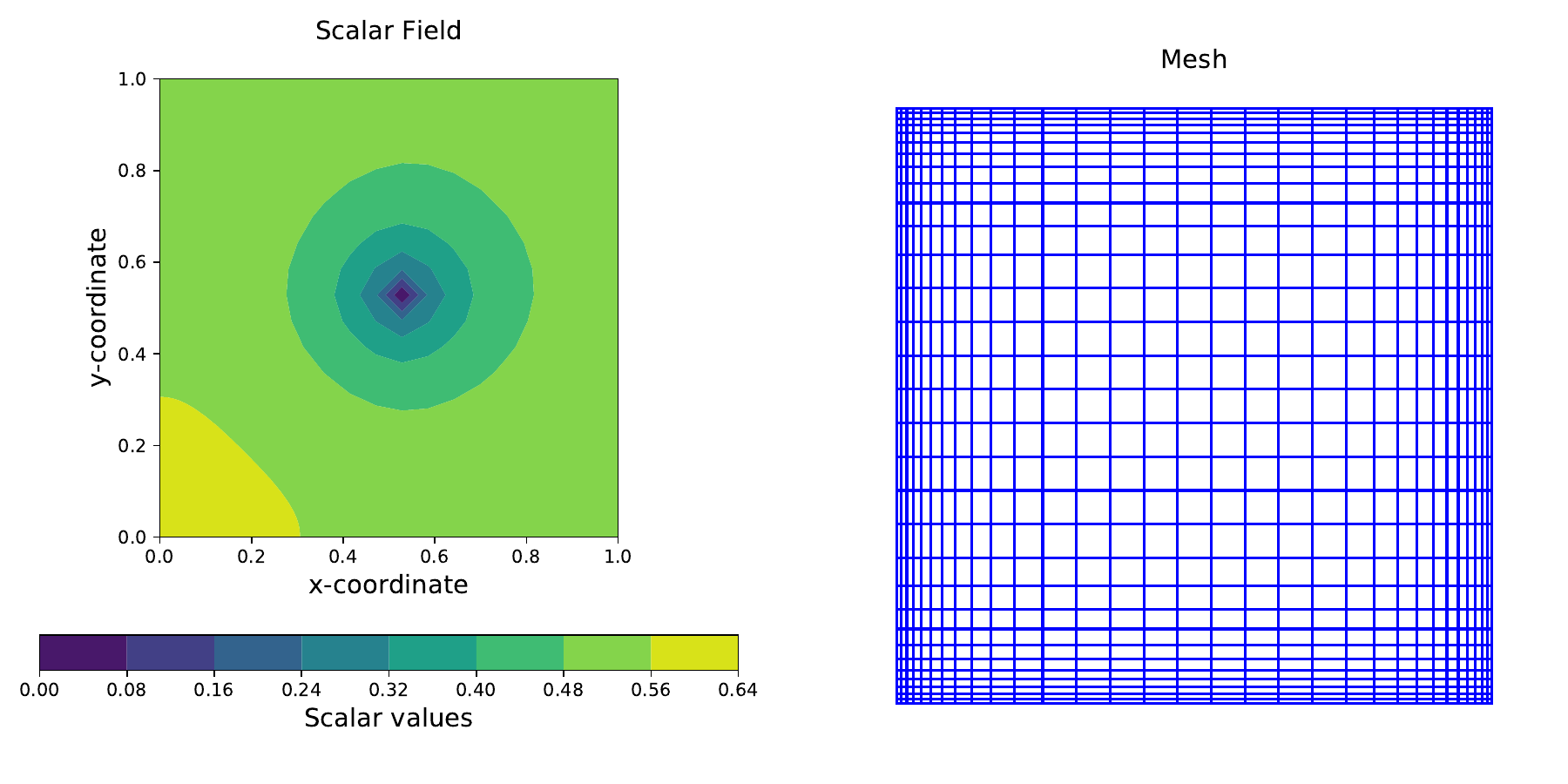}
  \caption{Solution and mesh for the solution of a 2D Laplacian with double
           Neumann boundary conditions on 32x32 mesh.}
  \label{fig-l2d_32x32_nnnn}
\end{figure}

Double Neumann conditions provide a further sophistication of 
pressure correction type equations. Here, the Laplacian is generated
on the same mesh as used for the momentum equations and, hence, 
is clustered next to boundary walls.
This is shown in \Cref{fig-l2d_32x32_nnnn} which is representative
of a wall bounded cavity.

At a solid wall, the normal gradient of pressure is zero and the
Neumann condition is used on all sides.
As with the double repeating boundary conditions, this leads to a degenerate
Laplacian and the same fix is used.

%
\subsubsection{Double Dirichlet}
\label{subsec-testcases-2d-DDDD}

\begin{figure}[ht]
  \centering
  \captionsetup{justification=centering}
  \includegraphics[clip, trim=1.0cm 1cm 1.0cm 0.4cm, width=0.66\textwidth]{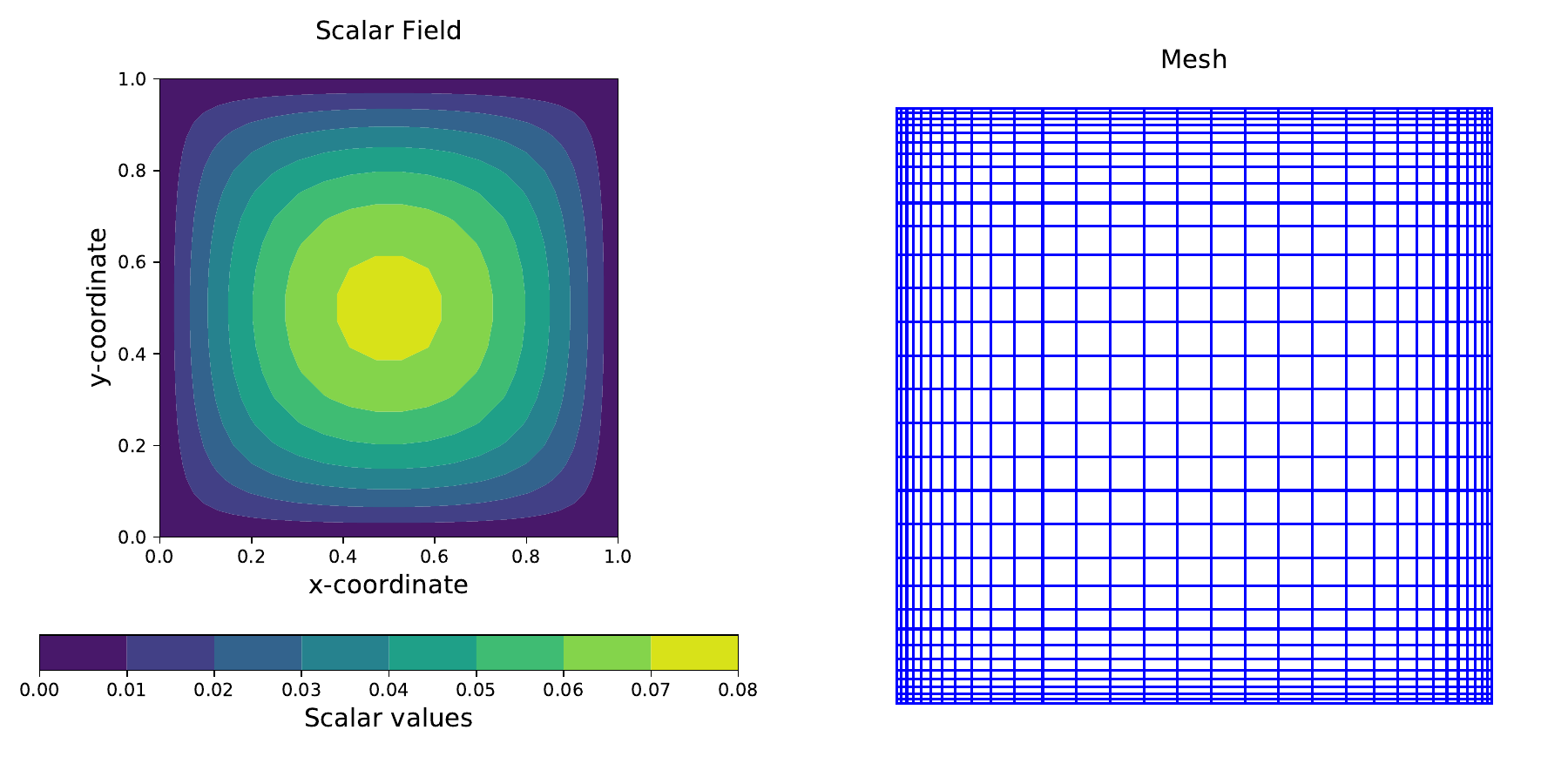}
  \caption{Solution and mesh for the solution of a 2D Laplacian with double
           Dirichlet boundary conditions on 32x32 mesh.}
  \label{fig-l2d_32x32_dddd}
\end{figure}

Double Dirichlet conditions provide a more standard Laplacian 
test case with all the boundary values set to a constant.
The same meshes as used as for the double Neumann case with 
clustering next to boundary walls.
This is shown in \Cref{fig-l2d_32x32_dddd} which is representative
of a wall bounded cavity.

%
\subsubsection{Wall bounded channel}
\label{subsec-testcases-2d-DNDD}
\begin{figure}[ht]
  \centering
  \captionsetup{justification=centering}
  \includegraphics[clip, trim=1.0cm 0cm 1.0cm 0.5cm, width=0.66\textwidth]{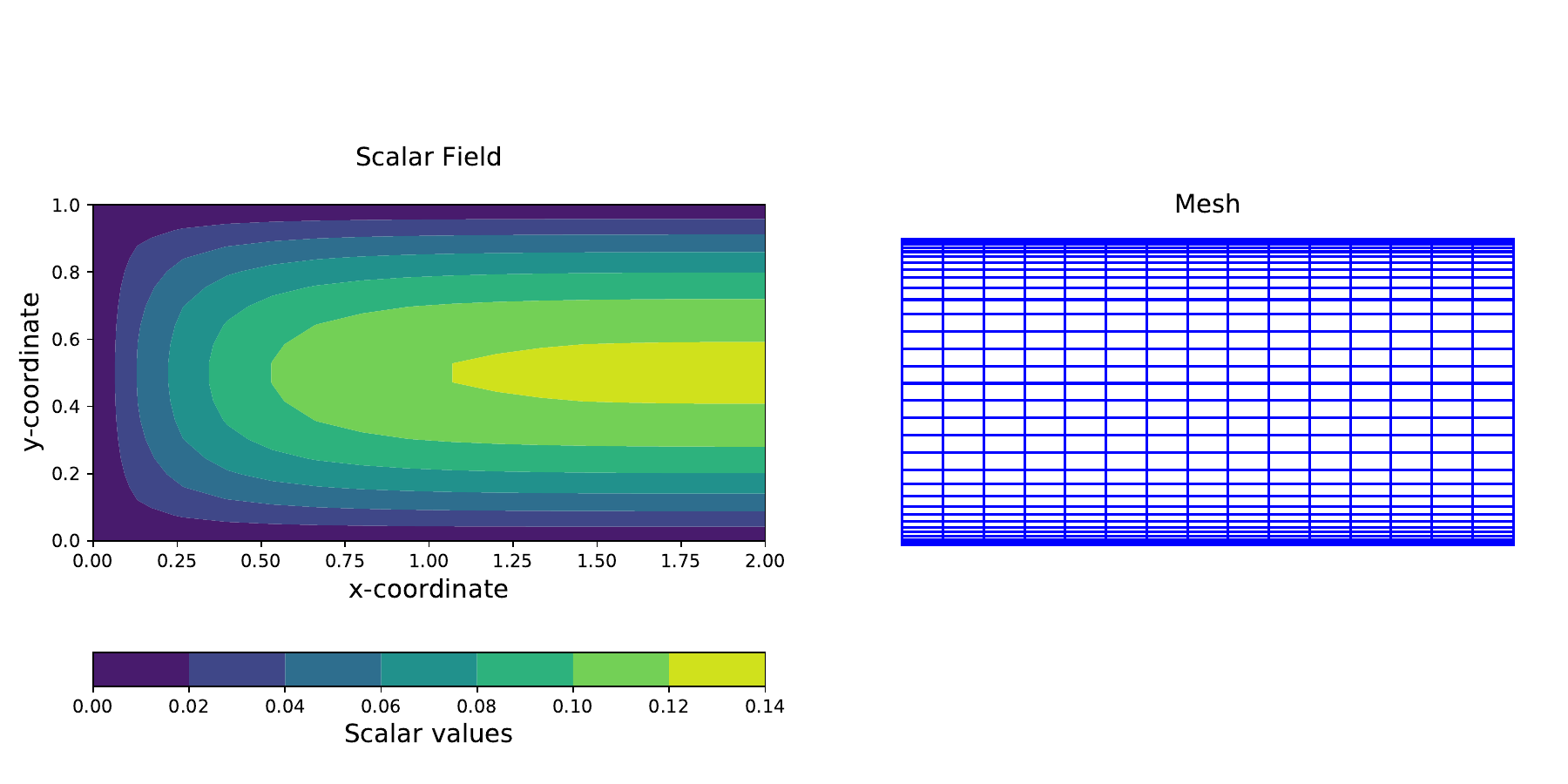}
  \caption{Solution and mesh for the solution of a 2D Laplacian for a wall
           bounded channel on a 16x32 mesh.}
  \label{fig-l2d_16x32_dndd}
\end{figure}

A wall bounded flow with a Dirichelet inlet and a Neumann outlet
is representative of Stokes flow in a channel.
Stokes flow ia a very low speed flow where the copnvective terms
in the Navier Stokes equation can be neglected.
As before, this test case is only indicative of the Laplacian for
the Navier Stokes stress tensor with constant viscosity.
Since this is a channel flow the length and number of points in
each direction are different as shown in \Cref{fig-l2d_16x32_dndd}.
Under the influence of a constant bulk force, the solution at the 
exit ($x=2$) approaches the fully developed Poiseuille flow profile.

%
\subsubsection{Test conditions}
\label{subsubsec_test_conds_2d}

The mesh parameters for the double repeat, Neumann and Dirichlet test cases are
shown in \Cref{tab-tests2d}. For the repeat conditions, setting $r=1.0$
ensures there is no clustering.

\begin{table}[h]
  \centering
  \captionsetup{justification=centering}
  \begin{tabular}{c c c  c c c c }
    \toprule
     Mesh  & $n_t$& $n_c$ & $r$ (Repeat)  & $r$ (Neumann/Dirichlet)  \\
    \midrule
     4x4   &  4   &   2   & 1.0  & 1.4  \\
     8x8   &  8   &   3   & 1.0  & 1.3  \\
     16x16 &  16  &   6   & 1.0  & 1.3  \\
     32x32 &  32  &  12   & 1.0  & 1.2  \\
     64x64 &  64   & 22   & 1.0  & 1.2  \\
    \bottomrule \\
  \end{tabular}
  \caption{Suggested test conditions for 2D Repeat and Neumann Laplacians.
           Clustering parameters are the same in both $x$ and $y$ directions.}
  \label{tab-tests2d}
\end{table}

\Cref{tab-tests2d_2} gives suggested mesh parameters for the wall bounded
flow Laplacians.

\begin{table}[h]
  \centering
  \begin{tabular}{c c c  c c c c }
    \toprule
        \multicolumn{1}{c}{} & \multicolumn{3}{c}{$x$-direction} & \multicolumn{3}{c}{$y$-direction} \\
     Mesh  & $n_t$& $n_c$ & $r$ & $n_t$& $n_c$ & $r$  \\
    \midrule
     4x8   &  4   &   2   & 1.0  & 8  &  4 &  1.3  \\
     8x16  &  8   &   3   & 1.0  & 16 &  6 &  1.3  \\
     16x32 &  16  &   6   & 1.0  & 32 & 12 &  1.2  \\
     32x64 &  32  &  12   & 1.0  & 64 & 22 & 1.2  \\
    \bottomrule \\
  \end{tabular}
  \caption{Suggested test conditions for 2D wall bounded Laplacians.}
  \label{tab-tests2d_2}
\end{table}

\begin{figure}[h]
  \centering
  \captionsetup{justification=centering}
  \includegraphics[width=0.66\textwidth]{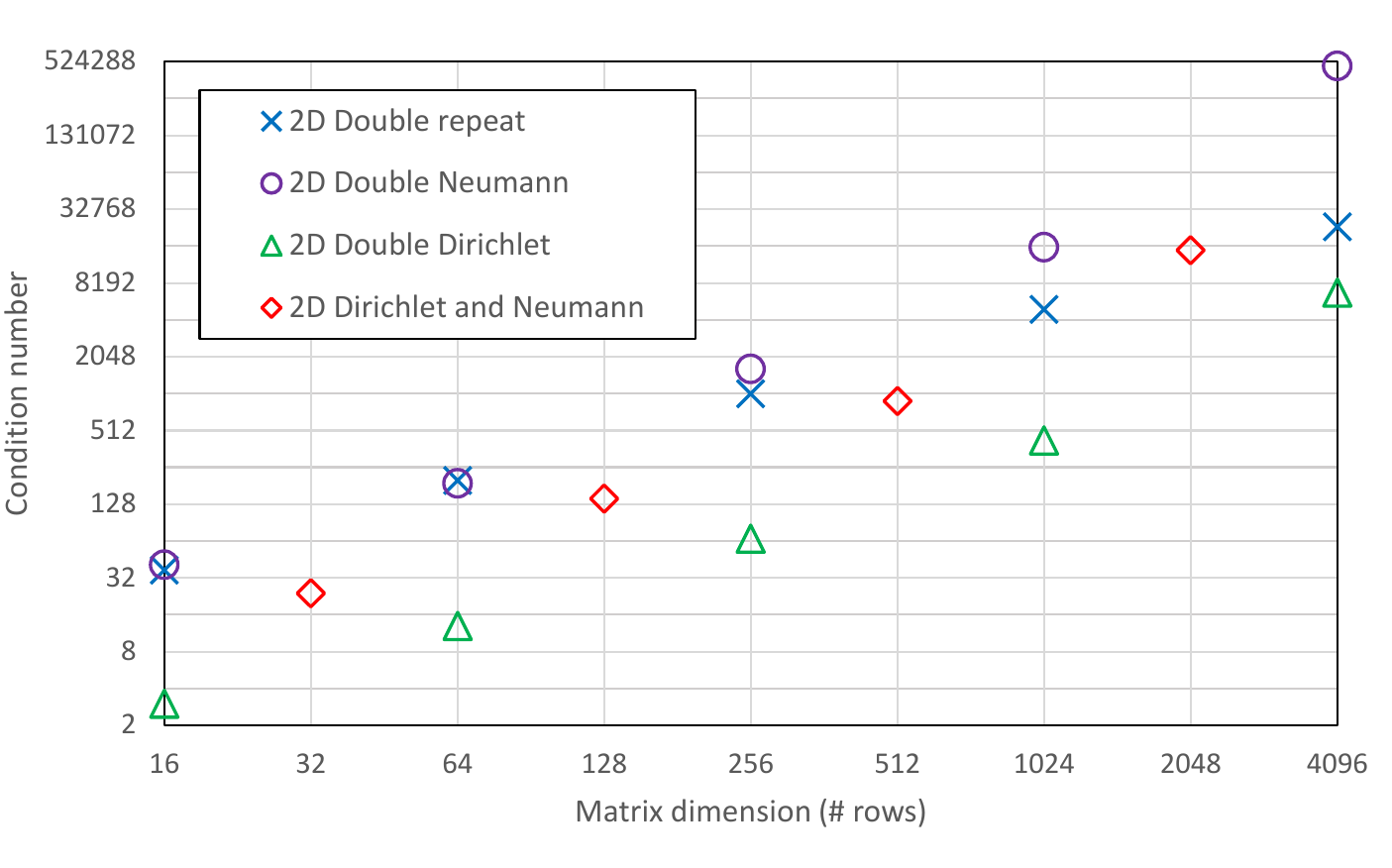}
  \caption{Condition numbers for the 2D Laplacians.}
  \label{fig-results-kappa}
\end{figure}

\Cref{fig-results-kappa} shows the condition numbers for all the
2D test cases. These range from $O(1)$ to $O(10^5)$.
For the smaller meshes, the double repeat and double Neumann cases
have similar condition numbers as the clustering has only a small effect
on the mesh. For the largest meshes the clustering has a larger effect
on the condition numbers. Note that the condition numbers are only available
for the non-degenerate Laplacians.

The condition numbers for the double Dirichlet cases are significantly lower than
the other cases and provide the best option for initial evaluations.

%
\subsection{3D test cases}
\label{subsec-testcases-3d}
Each of the 2D test cases can be easily extended into 3 dimensions.
Here, two channel flow cases are considered with meshes of
4x8x8 and 8x16x16. The Laplacian condition numbers are 23 and 93
respectively.

\Cref{tab-tests3d} gives suggested mesh parameters for the wall bounded
flow Laplacians.

\begin{table}[h]
  \centering
  \begin{tabular}{c c c c c c c c c c c}
    \toprule
        \multicolumn{1}{c}{} & \multicolumn{3}{c}{$x$-direction} & \multicolumn{3}{c}{$y$-direction}  & \multicolumn{3}{c}{$z$-direction} \\
     Mesh  & $n_t$& $n_c$ & $r$ & $n_t$& $n_c$ & $r$ & $n_t$& $n_c$ & $r$ \\
    \midrule
     4x8x8   &  4   &   2   & 1.0  & 8  &  3 &  1.3  & 8  &  3 &  1.3  \\
     8x16x16 &  8   &   3   & 1.0  & 16 &  6 &  1.3  & 16 &  6 &  1.3 \\

    \bottomrule \\
  \end{tabular}
  \caption{Suggested test conditions for 3D wall bounded Laplacians.}
  \label{tab-tests3d}
\end{table}

%% file: include/results.tex
%
\section{Sample results}
\label{sec-results}

To illustrate the use of L-QLES, a brief study has been performed
to analyse how different Laplacians affect the performance of two
matrix encoding formulations.

The first is the {\it Prepare-Select} encoding 
\cite{childs2012hamiltonian, babbush2018encoding, berry2015simulating, berry2018improved}
where the Laplacian is expressed as a Linear Combination of Unitaries (LCU).
Each unitary is typically a tensor product of 2x2 Pauli matrices, often
referred to as a Pauli string. Each Pauli string is multiplied by a coefficient
determined by the LCU decomposition. 
For real valued coefficients, the Laplacian, $L$ must be Hermitian.
If it is not Hermitian, the LCU is applied to the bipartite matrix:

\begin{equation}
  \begin{pmatrix}
    0   & L \\
    L^* & 0 \\
  \end{pmatrix}
\label{eqn-lcu}
\end{equation}

This requires an additional qubit in the select register.

The second encoding is the {\it FABLE} method \cite{camps2022fable}. 
FABLE is derived from the matrix encoding using a row-column oracle 
\cite{lin2022lecture} which creates multiplexed rotations on an 
ancilla qubit for each entry in the matrix.
FABLE uses Gray code ordering to
convert the multiplexed rotations into an interleaving sequence of
uncontrolled rotations and CNOT gates
\cite{mottonen2004transformation}. 
A linear system is solved to obtain the uncontrolled rotation
angles from the original angles. Under certain circumstances a
large number of these angles are zero, or close to zero, and the FABLE
circuit is far shallower that the oracle based circuit.
FABLE can encode non-Hermitian matrices.

In order to compare the performance of Pauli (Prep-Select) and FABLE a means
of normalising their outputs is needed. For Prep-Select the number of
non-zero entries in the Laplacian is used.
For FABLE, the total number of entries in the matrix ($4^n$) is used as the
row-column oracle creates a multiplexed rotation for every matrix position.
In the following analysis, only the rotation gates in the FABLE
encoding are used in the comparisons.

Note that all the FABLE calculations were run with a angle tolerance of
$10^{-9}$. This is to eliminate rounding errors in the double precision 
Python matrix assembly calculations. Effectively the \textit{approximation}
part of the FABLE algorithm is not being used.
Similarly, the LCU decomposition into Pauli strings can be subject to
rounding error and strings with coefficients less than $10^{-9}$ are
also ignored.
Changing the tolerances will undoubtedly affect the results below but
that is beyond the scope of this work which is to demonstrate the uses to
which the testing framework can be put.
The impact of changing the tolerance is not just the effect on circuit
depth but the fidelity of the QLES solution.

%
\subsection{1D Dirichlet and repeating}
\label{subsec-results-1d}

\begin{figure}[h]
  \centering
  \captionsetup{justification=centering}
  \includegraphics[width=0.66\textwidth]{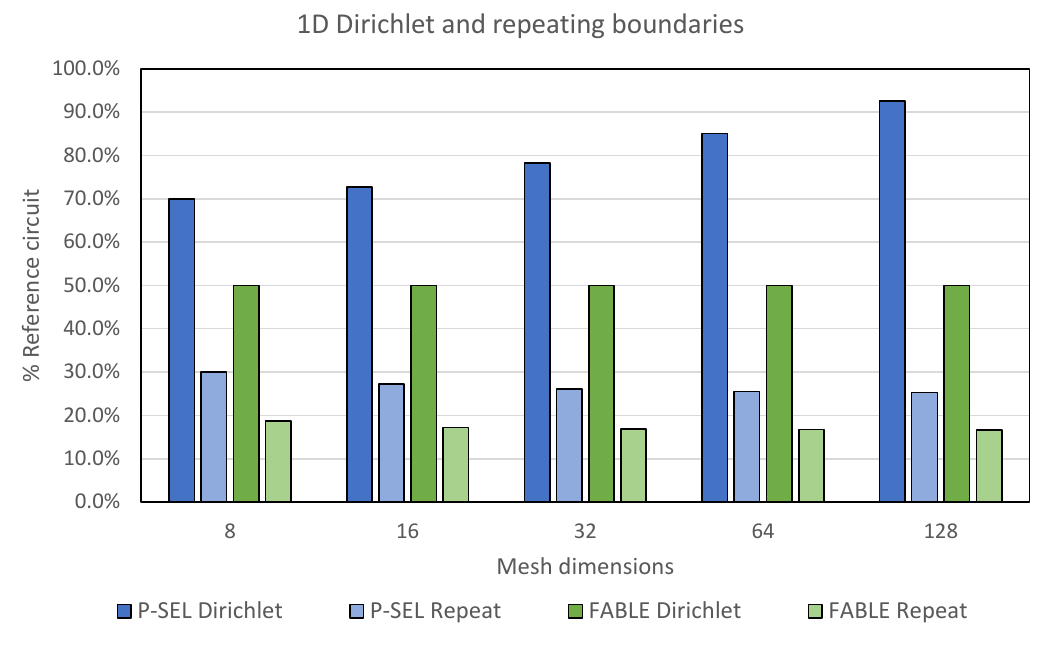}
  \caption{Normalised circuit counts for 1D Laplacian with Dirichlet and repeat cases.}
  \label{fig-results-1D}
\end{figure}

\Cref{fig-results-1D} compares the normalised Pauli and FABLE encoding for the
1d Laplacian with Dirichlet and repeating boundary conditions. 
The latter uses the degenerate version of the Laplacian.
Both FABLE results (dark and light green) match those reported by \cite{camps2022fable}.

%
\subsection{2D Double repeating}
\label{subsec-results-2d-RRRR}

\begin{figure}[h]
  \centering
  \captionsetup{justification=centering}
  \includegraphics[width=0.66\textwidth]{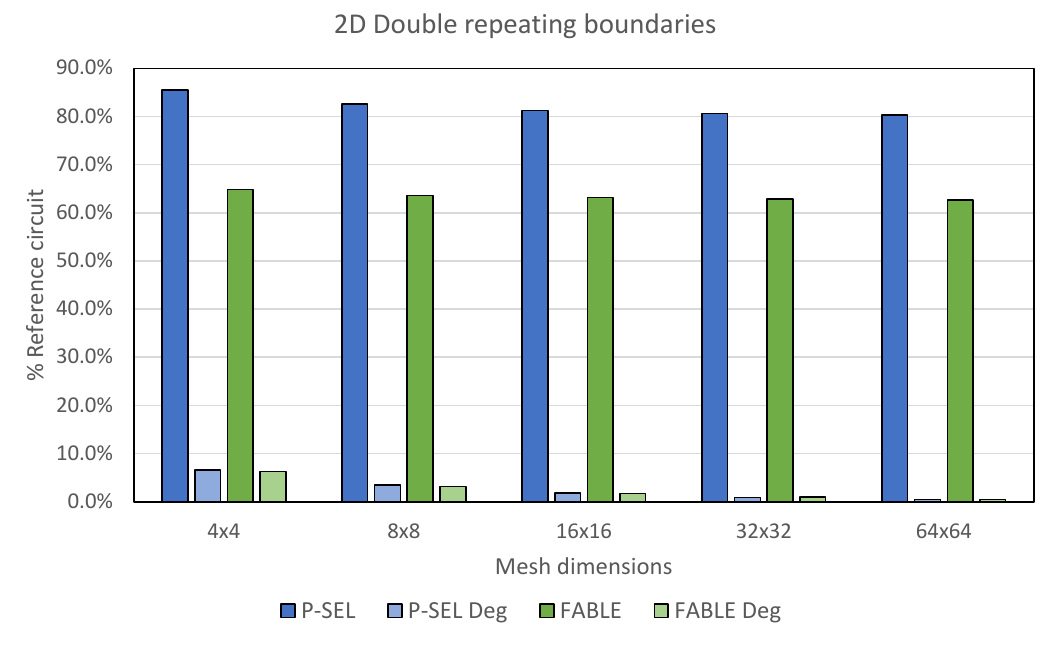}
  \caption{Normalised circuit counts for 2D double repeat case with and without degeneracy.}
  \label{fig-results-TGV}
\end{figure}

\Cref{fig-results-TGV} compares the normalised Pauli and FABLE encoding for the
2d double repeating case with degenerate and non-degenerate Laplacians.
The degenerate FABLE results (light green) match those for the 4x4 and 8x8
Laplacians reported by \cite{camps2022fable}, 
as expected since the Laplacians are identical.
The Prep-Select encoding shows almost identical behaviour to FABLE for the 
degenerate Laplacian. For the 64x64 mesh, both encodings have a normalised
count of 0.5\%.

However, when the degeneracy is removed, the behaviour is quite different.
Prep-Select encoding (dark blue) has a count of around 80\% and FABLE has
a count of around 60\% for all meshes. It is striking that a modification
to one row can cause such a significant change to both schemes.

\begin{table}[h]
  \centering
  \begin{tabular}{c c c  c c c c }
    \toprule
        \multicolumn{1}{c}{} & \multicolumn{2}{c}{Prep-Select} & \multicolumn{2}{c}{FABLE} \\
     Mesh  & non-deg & deg & non-deg & deg  \\
    \midrule
     4x4   &  12  &   8   & 9   & 9   \\
     8x8   &  16  &  11   & 13  & 13  \\
     16x16 &  20  &  14   & 17  & 17  \\
     32x32 &  24  &  17   & 21  & 21  \\
     64x64 &  28  &  20   & 25  & 25  \\
    \bottomrule \\
  \end{tabular}
  \caption{Total number of qubits used to encode for 2D double repeat Laplacians.}
  \label{tab-rrrr-nq}
\end{table}

\Cref{tab-rrrr-nq} shows the total number of qubits needed by each encoder.
Since FABLE processes all the elements in the Laplacian, the number of qubits
is independent of the gate count.
In contrast, the Prep operator is a state loader for the LCU coefficients and 
the number of qubits needed depends on the number of coefficients to be loaded.

%
\subsection{2D Double Neumann}
\label{subsec-results-2d-NNNN}

\begin{figure}[h]
  \centering
  \captionsetup{justification=centering}
  \includegraphics[width=0.66\textwidth]{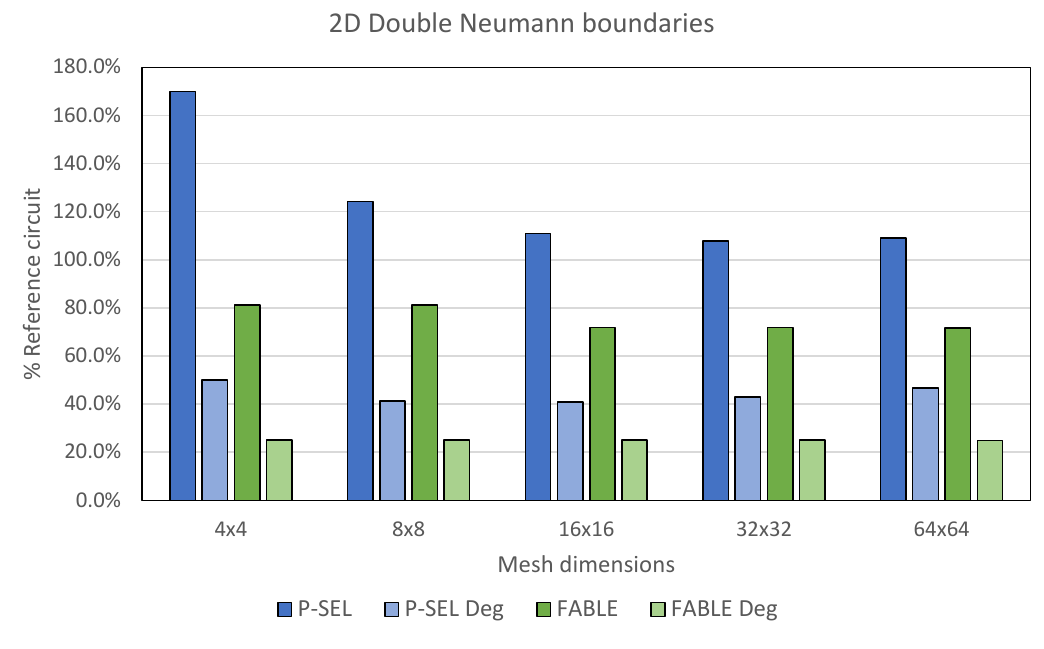}
  \caption{Normalised circuit counts for 2D double Neumann case with and without degeneracy.}
  \label{fig-results-PC}
\end{figure}

\Cref{fig-results-PC} compares the normalised Pauli and FABLE encoding for the
2d double Neumann case with degenerate and non-degenerate Laplacians.
The behaviour is quite different to the double repeat case.
Two changes have been made: the boundary condition and the clustering of the mesh.
Testing the 16x16 mesh with a uniform mesh gave identical results 
for FABLE and for Prep-Select the count percentage reduced from
111\% to 102\%. 
The boundary condition is having the largest effect.

Taking the 4x4 mesh, the degenerate double repeat Laplacian, effectively, has 16 interior
nodes, one of which is updated to remove the degeneracy.
The double Neumann mesh has 4 interior nodes and 12 boundary nodes with one of the
interior nodes being updated to remove the degeneracy. This gives a far more irregular
sparsity pattern for the encoders to exploit.
The different nature of the interior and boundary equations also means the degenerate
Laplacians do not achieve the same reductions as the double repeat cases.
It is surprising that the effect of the irregularity does not diminish.

\begin{table}[h]
  \centering
  \begin{tabular}{c c c  c c c c }
    \toprule
        \multicolumn{1}{c}{} & \multicolumn{2}{c}{Prep-Select} & \multicolumn{2}{c}{FABLE} \\
     Mesh  & non-deg & deg & non-deg & deg  \\
    \midrule
     4x4   &  12  &  10   & 9   & 9   \\
     8x8   &  16  &  14   & 13  & 13  \\
     16x16 &  20  &  18   & 17  & 17  \\
     32x32 &  24  &  22   & 21  & 21  \\
     64x64 &  28  &  27   & 25  & 25  \\
    \bottomrule \\
  \end{tabular}
  \caption{Total number of qubits used to encode for 2D double Neumann Laplacians.}
  \label{tab-nnnn-nq}
\end{table}

\Cref{tab-nnnn-nq} shows the total number of qubits needed by each encoder. 
\Cref{subsec-testcases-2d-NNNN} described this case as representative of a CFD
pressure correction equation. In staggered-grid pressure correction formulations
\cite{versteeg2007introduction}, the Neumann boundary condition becomes implicit
and there are no boundary nodes in the Laplacian.

%
\subsection{2D Double Dirichlet}
\label{subsec-results-2d-DDDD}

\begin{figure}[h]
  \centering
  \captionsetup{justification=centering}
  \includegraphics[width=0.66\textwidth]{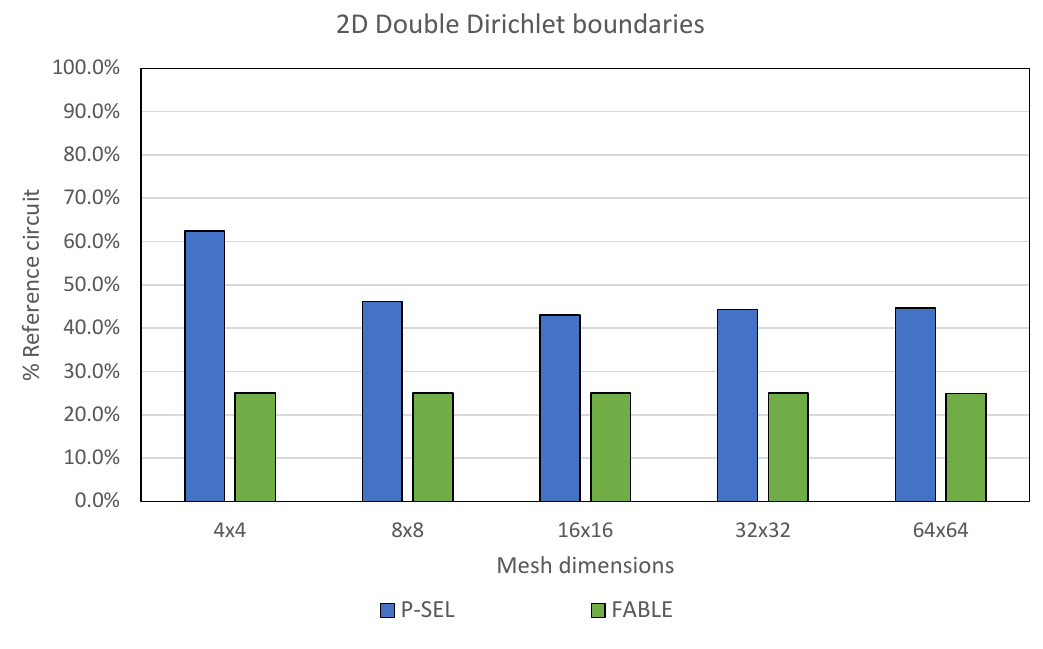}
  \caption{Normalised circuit counts for 2D double Dirichlet case.}
  \label{fig-results-cavity}
\end{figure}

Given the difference in behaviour between the double repeat and the double Neumann
cases, the clustered Neumann meshes were also analysed using Dirichlet conditions
on all boundaries. There is no degeneracy to be removed in this case.
\Cref{fig-results-PC} compares the normalised Pauli and FABLE encoding for the
2D double Dirichlet cases.

For Prep-Select, the number of Pauli strings in the LCU was identical to those
in \Cref{fig-results-PC}. The FABLE counts were almost identical, with differences
of less than 0.1\%.

%
\subsection{2D Wall bounded channel}
\label{subsec-results-2d-DNDD}

\begin{figure}[h]
  \centering
  \captionsetup{justification=centering}
  \includegraphics[width=0.66\textwidth]{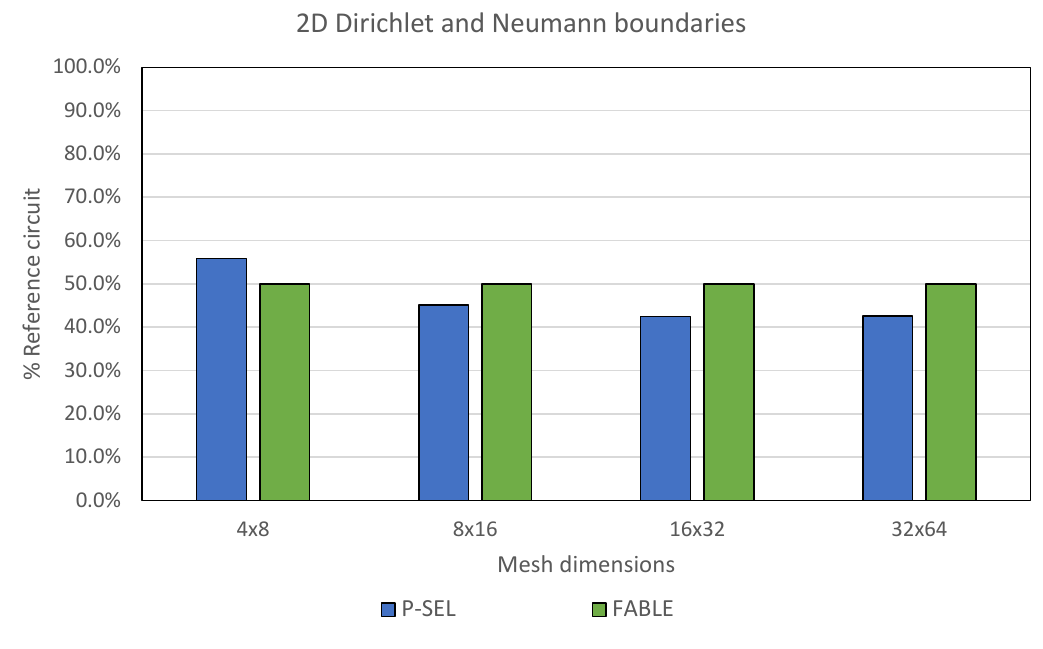}
  \caption{Normalised circuit counts for 2D wall bounded channel.}
  \label{fig-results-Stokes}
\end{figure}

\Cref{fig-results-Stokes} compares the normalised Pauli and FABLE encoding for the wall
bounded channel. 
The FABLE results consistently have a 50\% count for all the cases.
There is some variation in the Prep-Select counts which settles to around 40\%
for the higher meshes.

\begin{table}[h]
  \centering
  \begin{tabular}{c c c c c c c }
    \toprule
     Mesh  & Prep-Select & FABLE \\
    \midrule
     4x8   &  12  & 11  \\
     8x16  &  16  & 15  \\
     16x32 &  20  & 19  \\
     32x64 &  24  & 23  \\
    \bottomrule \\
  \end{tabular}
  \caption{Total number of qubits used to encode for 2D wall bounded Laplacians.}
  \label{tab-dndd-nq}
\end{table}

\Cref{tab-dndd-nq} shows the total number of qubits needed by each encoder. 
The difference between Prep-Select and FABLE is only 1 qubit, which is lower
than previous cases where the non-degenerate cases differed by 3 qubits.

%
\subsection{3D Wall bounded channel}
\label{subsec-results-3d-DNDDDD}

\begin{figure}[h]
  \centering
  \captionsetup{justification=centering}
    \begin{subfigure}{0.45\textwidth}
    \includegraphics[clip, trim=1.0cm 0cm 2.0cm 1cm, width=\textwidth]{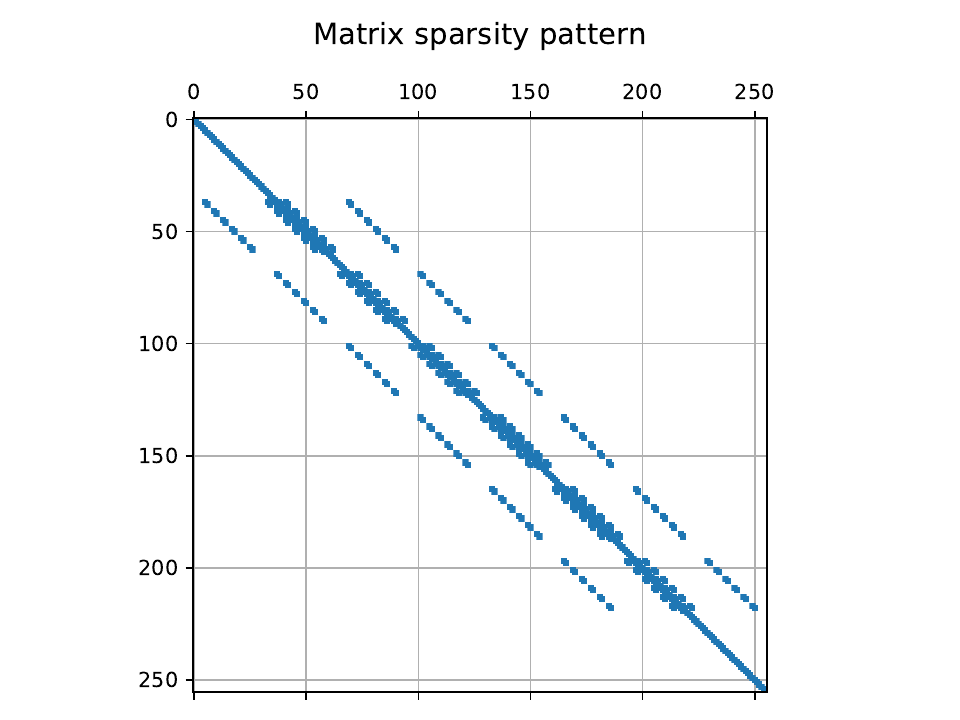}
    \caption{Coordinate based ordering.}
    \label{fig-l3d_4x8x8_pat}
  \end{subfigure}
  \captionsetup{justification=centering}
    \begin{subfigure}{0.45\textwidth}
    \includegraphics[clip, trim=1.0cm 0cm 2.0cm 1cm, width=\textwidth]{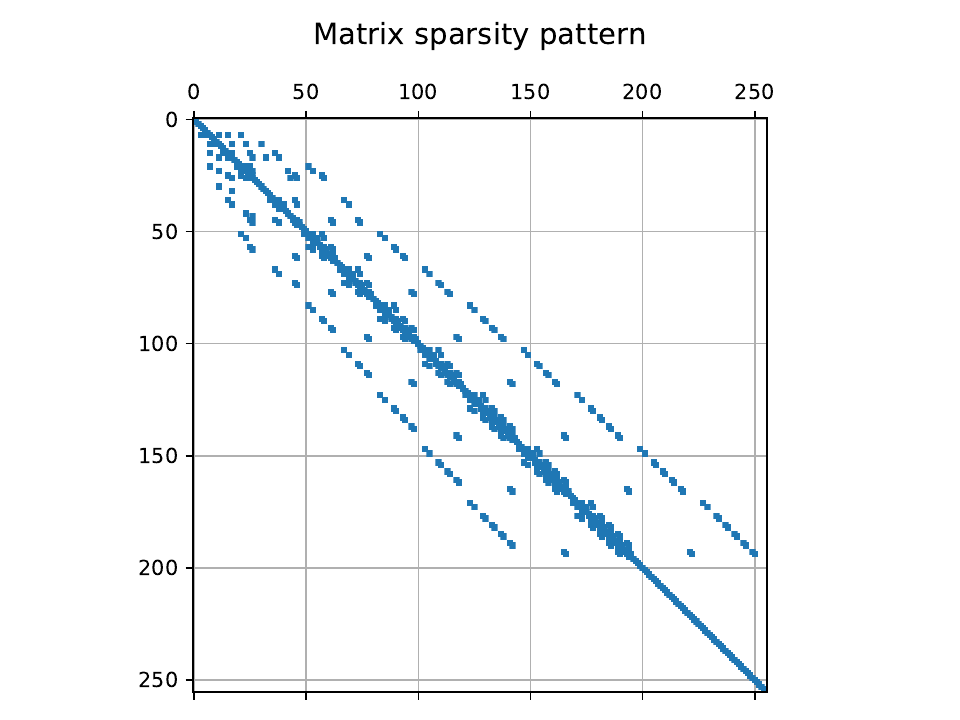}
    \caption{Shell based ordering.}
    \label{fig-l3d_4x8x8_r_pat}
  \end{subfigure}
  \caption{Sparsity patterns for 4x8x8 3D Laplacian.}
  \label{fig-l3d_4x8x8}
\end{figure}

The 3D test case is used to demonstrate the effect of reordering on each encoder.
\Cref{fig-l3d_4x8x8} shows the sparsity patterns of the original and reordered
Laplacians.

\begin{table}[h]
  \centering
  \begin{tabular}{c c c c c c c c c}
    \toprule
    \multicolumn{3}{c}{} & \multicolumn{3}{c}{Original} & \multicolumn{3}{c}{Reordered} \\
     Mesh  & $\kappa$ & non-zeros & Pauli strings & count & qubits & Pauli strings & count & qubits \\
    \midrule
     4x8x8   & 22.8 &  724  & 272  & 37.6\% & 18 & 8384 & 1158\%&  23  \\
     8x16x16 & 93.1 &  9300 & 2496 & 26.8\% & 24 &   -  &  -    &  -   \\
    \bottomrule \\
  \end{tabular}
  \caption{LCU and qubit counts for Prep-Select encoding for 3D wall bunded flows.}
  \label{tab-dndddd-lcu}
\end{table}

\Cref{tab-dndddd-lcu} shows the number of Pauli strings and qubits for the original
and reordered Laplacians.
Firstly, the Pauli string counts are lower than for the 2D cases.
However, the reordering has a huge effect, increasing the number of Pauli strings
by over a factor of over 30 for the 4x8x8 mesh.
Processing of the 8x16x16 mesh was not attempted.

In contrast the FABLE counts for the original and reordered Laplacians were
25\% and 100\% for both cases. Whilst the row-column oracle produces the same set
of rotation angles, they are in a different order. This in turn changes the derived
angles and/or their Gray code order.

%% file: include/close.tex
%
\section{Closing remarks}
\label{sec-close}

A test case framework, L-QLES, has been presented which contains important
features of industrial applications of sparse matrix algebra.
Sample test cases have demonstrated the flexibility to investigate features 
of quantum algorithms, in this case matrix encoding.
With some differences in degree, all the 2D cases show asymptotic behaviour
as the mesh dimensions increase.
This suggests progress on encoding schemes can be made on smaller meshes.
Other aspects of QLES have not been investigated but the rate of increase
of condition number with mesh size demonstrates the need to consider industrially
representative cases.

An important part of any application of QLES is the classical preprocessing needed
to prepare the quantum circuit. L-QLES also allows aspects such as matrix encoding to
be evaluated.
For example, the FABLE encoding of the 64x64 mesh took 20s, whereas the Prep-Select
encoding took 6,639 seconds (1:50:33).
Both were run on an
Intel\textsuperscript{\tiny\textregistered}
Core\textsuperscript{\tiny\textcopyright} i9 12900K 3.2GHz
Alder Lake 16 core processor with 64GB of 3,200MHz DDR4 RAM.

%
\section{Data availability}
\label{sec-data}
The L-QLES scripts and sample input files are available from
\href{https://github.com/rolls-royce/qc-cfd/tree/main/L-QLES}{https://github.com/rolls-royce/qc-cfd/tree/main/L-QLES}.

%
\section{Acknowledgements}
The idea for this test case framework arose in discussions with 
Christoph S{\"u}nderhauf and Joan Camps at their Riverlane offices and 
I am very grateful to them for their inspiration.
I hope L-QLES meets our goal of allowing quantum algorithm developers to stress
test their algorithms without the need for esoteric domain knowledge or reliance
on matrices supplied by industry. 
I would also like to thank Jarrett Smalley and Tony Phipps 
of Rolls-Royce and Philippa Rubin of the STFC Hartree Centre for their helpful 
comments on this work.

The permission of Rolls-Royce to publish this work is gratefully acknowledged.
This work was completed under funding received under the UK's
Commercialising Quantum Technologies Programme (Grant reference 10004857).

%% file: include/appendix-ksum.tex
%
\section{Kronecker sums vs direct discretisation}
\label{sec-ksum-vs-disc}

For Cartesian lattice based meshes, a convenient way
to create a 2D Laplacian operator, $L$ from separate 
1D operators, $L_{xx}$ and $L_{yy}$ is using the
Kronecker sum:

\begin{equation}
    L = L_{yy} \oplus L_{xx} = I \otimes L_{xx} + L_{yy} \otimes I
    \label{eqn_2dlap_oper}
\end{equation}

The ordering of the sum depends on the indexing of the nodes
in the resulting 2D lattice mesh.
\Cref{fig-2dlattice} shows a 4x4 lattice. 
In the formulation of \Cref{eqn_2dlap_oper}, the $L_{xx}$
operator acts in the horizontal direction and $L_{yy}$ in
the vertical direction.

\begin{figure}[h!]
  \centering
  \begin{tikzpicture}[]
    \draw [step=1.5cm,] (0,0) grid (4.5,4.5);
      \begin{scope}[node distance=0.1mm and 0.1mm]
        \node (b) at (0,0) {};
        \node [below left=of b, xshift=0.1cm, yshift=0.1cm]{0};
        \node [below left=of b, xshift=1.6cm, yshift=0.1cm]{1};
        \node [below left=of b, xshift=3.1cm, yshift=0.1cm]{2};
        \node [below left=of b, xshift=4.6cm, yshift=0.1cm]{3};
        \node [below left=of b, xshift=0.1cm, yshift=1.6cm]{4};
        \node [below left=of b, xshift=1.6cm, yshift=1.6cm]{5};
        \node [below left=of b, xshift=3.1cm, yshift=1.6cm]{6};
        \node [below left=of b, xshift=4.6cm, yshift=1.6cm]{7};
        \node [below left=of b, xshift=0.1cm, yshift=3.1cm]{8};
        \node [below left=of b, xshift=1.6cm, yshift=3.1cm]{9};
        \node [below left=of b, xshift=3.1cm, yshift=3.1cm]{10};
        \node [below left=of b, xshift=4.6cm, yshift=3.1cm]{11};
        \node [below left=of b, xshift=0.1cm, yshift=4.6cm]{12};
        \node [below left=of b, xshift=1.6cm, yshift=4.6cm]{13};
        \node [below left=of b, xshift=3.1cm, yshift=4.6cm]{14};
        \node [below left=of b, xshift=4.6cm, yshift=4.6cm]{15};
      \end{scope}
  \end{tikzpicture}
  \caption{Indexing of nodes on a 4x4 Cartesian lattice mesh.}  
  \label{fig-2dlattice}
\end{figure}

The following subsections analyse how well the construction
in \Cref{eqn_2dlap_oper} preserves the relevant boundary conditions.

%
\subsection{Repeat-Repeat boundary conditions}
\label{subsec-rep-rep}
For a uniform mesh, the 1D Laplacian operator with repeating boundary conditions is: 

\begin{equation}
  L_{xx} = L_{yy} =
    \begin{pmatrix*}[r]
     a & -b &  . & -b \\
    -b &  a & -b &  . \\
     . & -b &  a & -b \\
    -b &  . & -b &  a \\
  \end{pmatrix*}
  \label{eqn-lxx-rep}
\end{equation}

Many analyses use $a=1$ and $b=0.5$, but this does not have to be the case. 
Note that this is not the usual CFD formulation where the values on opposite
sides of the repeating boundary are equal.
This formulation can be envisaged by taking the bottom line of the lattice
in \Cref{fig-2dlattice} and placing a nodes at positions (-1, 0) and (4,0) 
with the same uniform spacing as the rest of the lattice.
Here, the indices refer to (x,y) coordinate positions.
The repeating boundary condition sets, $\phi_{3,0} = \phi_{-1,0}$ and
$\phi_{0,0} = \phi_{4,0}$.
This is an important distinction as the two formulations have different
periods.

If repeating conditions are applied in both the $x$ and
$y$ directions, the 2D Laplacian constructed by using \Cref{eqn_2dlap_oper} is:

\setcounter{MaxMatrixCols}{16}
\begin{equation}
  L\ket{\phi} = \\
  \begin{pmatrix}
    2a & -b &  . & -b & -b &  . &  . &  . &  . &  . &  . &  . & -b &  . &  . &  . \\  
    -b & 2a & -b &  . &  . & -b &  . &  . &  . &  . &  . &  . &  . & -b &  . &  . \\  
     . & -b & 2a & -b &  . &  . & -b &  . &  . &  . &  . &  . &  . &  . & -b &  . \\  
    -b &  . & -b & 2a &  . &  . &  . & -b &  . &  . &  . &  . &  . &  . &  . & -b \\  
    -b &  . &  . &  . & 2a & -b &  . & -b & -b &  . &  . &  . &  . &  . &  . &  . \\  
     . & -b &  . &  . & -b & 2a & -b &  . &  . & -b &  . &  . &  . &  . &  . &  . \\  
     . &  . & -b &  . &  . & -b & 2a & -b &  . &  . & -b &  . &  . &  . &  . &  . \\  
     . &  . &  . & -b & -b &  . & -b & 2a &  . &  . &  . & -b &  . &  . &  . &  . \\  
     . &  . &  . &  . & -b &  . &  . &  . & 2a & -b &  . & -b & -b &  . &  . &  . \\  
     . &  . &  . &  . &  . & -b &  . &  . & -b & 2a & -b &  . &  . & -b &  . &  . \\  
     . &  . &  . &  . &  . &  . & -b &  . &  . & -b & 2a & -b &  . &  . & -b &  . \\  
     . &  . &  . &  . &  . &  . &  . & -b & -b &  . & -b & 2a &  . &  . &  . & -b \\  
    -b &  . &  . &  . &  . &  . &  . &  . & -b &  . &  . &  . & 2a & -b &  . & -b \\  
     . & -b &  . &  . &  . &  . &  . &  . &  . & -b &  . &  . & -b & 2a & -b &  . \\  
     . &  . & -b &  . &  . &  . &  . &  . &  . &  . & -b &  . &  . & -b & 2a & -b \\  
     . &  . &  . & -b &  . &  . &  . &  . &  . &  . &  . & -b & -b &  . & -b & 2a \\  
  \end{pmatrix}
  \begin{pmatrix}
  \phi_{0}  \\ \phi_{1}  \\ \phi_{2}  \\ \phi_{3}  \\ \phi_{4}  \\ \phi_{5} \\
  \phi_{6}  \\ \phi_{7}  \\ \phi_{8}  \\ \phi_{9}  \\ \phi_{10} \\ \phi_{11} \\
  \phi_{12} \\ \phi_{13} \\ \phi_{14} \\ \phi_{15} \\
  \end{pmatrix}
  \label{eqn_lap2d_repx2}
\end{equation}

The matrix is shown acting on a vector to aid visual inspection that the correct
2D Laplacian has been constructed.
Doubly repeating boundary conditions such as this are relevant to CFD for
test cases such as the Taylor-Green Vortex \cite{orszag1974numerical}
and the decay of isotropic turbulence \cite{comte1971simple}.

%
\subsection{Dirichlet-Dirichlet boundary conditions}
\label{subsec-dir-dir}
For a uniform mesh, the 1D Laplacian operator with Dirichlet boundary conditions
at each end is: 

\begin{equation}
  L_{xx} = L_{yy} =
    \begin{pmatrix*}[r]
     a &  . &  . &  . \\
    -b &  a & -b &  . \\
     . & -b &  a & -b \\
     . &  . &  . &  a \\
  \end{pmatrix*}
  \label{eqn-lxx-dir}
\end{equation}

If Dirichlet boundary conditions are applied in both the $x$ and
$y$ directions, the 2D Laplacian constructed by using \Cref{eqn_2dlap_oper} is:

\setcounter{MaxMatrixCols}{16}
\begin{equation}
  L\ket{\phi} = \\
  \begin{pmatrix}
    2a &  . &  . &  . &  . &  . &  . &  . &  . &  . &  . &  . &  . &  . &  . &  . \\  
    -b & 2a & -b &  . &  . &  . &  . &  . &  . &  . &  . &  . &  . &  . &  . &  . \\  
     . & -b & 2a & -b &  . &  . &  . &  . &  . &  . &  . &  . &  . &  . &  . &  . \\  
     . &  . &  . & 2a &  . &  . &  . &  . &  . &  . &  . &  . &  . &  . &  . &  . \\  
    -b &  . &  . &  . & 2a &  . &  . &  . & -b &  . &  . &  . &  . &  . &  . &  . \\  
     . & -b &  . &  . & -b & 2a & -b &  . &  . & -b &  . &  . &  . &  . &  . &  . \\  
     . &  . & -b &  . &  . & -b & 2a & -b &  . &  . & -b &  . &  . &  . &  . &  . \\  
     . &  . &  . & -b &  . &  . &  . & 2a &  . &  . &  . & -b &  . &  . &  . &  . \\  
     . &  . &  . &  . & -b &  . &  . &  . & 2a &  . &  . &  . & -b &  . &  . &  . \\  
     . &  . &  . &  . &  . & -b &  . &  . & -b & 2a & -b &  . &  . & -b &  . &  . \\  
     . &  . &  . &  . &  . &  . & -b &  . &  . & -b & 2a & -b &  . &  . & -b &  . \\  
     . &  . &  . &  . &  . &  . &  . & -b &  . &  . &  . & 2a &  . &  . &  . & -b \\  
     . &  . &  . &  . &  . &  . &  . &  . &  . &  . &  . &  . & 2a &  . &  . &  . \\  
     . &  . &  . &  . &  . &  . &  . &  . &  . &  . &  . &  . & -b & 2a & -b &  . \\  
     . &  . &  . &  . &  . &  . &  . &  . &  . &  . &  . &  . &  . & -b & 2a & -b \\  
     . &  . &  . &  . &  . &  . &  . &  . &  . &  . &  . &  . &  . &  . &  . & 2a \\  
  \end{pmatrix}
  \begin{pmatrix}
  \phi_{0}  \\ \phi_{1}  \\ \phi_{2}  \\ \phi_{3}  \\ \phi_{4}  \\ \phi_{5} \\
  \phi_{6}  \\ \phi_{7}  \\ \phi_{8}  \\ \phi_{9}  \\ \phi_{10} \\ \phi_{11} \\
  \phi_{12} \\ \phi_{13} \\ \phi_{14} \\ \phi_{15} \\
  \end{pmatrix}
  \label{eqn_lap2d_dirx2}
\end{equation}

From inspection of \Cref{eqn_lap2d_dirx2}, we can see that the corner nodes 0, 3, 12 and 15
have the correct boundary condition; and, the interior nodes 5, 6, 9 and 10 have the correct
Laplacian operator. The remaining boundary nodes 1, 2, 4, 7, 8, 11, 13 and 14 do not
have the Dirichlet boundary condition correctly applied.
Each of these rows in the matrix should have a single entry $2a$ on the diagonal, as shown in
\Cref{eqn_lap2d_dirx2_bc}. This matrix has few off diagonal entries due to the small
ratio of internal nodes to boundary nodes. For larger meshes, the number of internal nodes
is dominant.

\setcounter{MaxMatrixCols}{16}
\begin{equation}
  L\ket{\phi} = \\
  \begin{pmatrix}
    2a &  . &  . &  . &  . &  . &  . &  . &  . &  . &  . &  . &  . &  . &  . &  . \\  
     . & 2a &  . &  . &  . &  . &  . &  . &  . &  . &  . &  . &  . &  . &  . &  . \\  
     . &  . & 2a &  . &  . &  . &  . &  . &  . &  . &  . &  . &  . &  . &  . &  . \\  
     . &  . &  . & 2a &  . &  . &  . &  . &  . &  . &  . &  . &  . &  . &  . &  . \\  
     . &  . &  . &  . & 2a &  . &  . &  . &  . &  . &  . &  . &  . &  . &  . &  . \\  
     . & -b &  . &  . & -b & 2a & -b &  . &  . & -b &  . &  . &  . &  . &  . &  . \\  
     . &  . & -b &  . &  . & -b & 2a & -b &  . &  . & -b &  . &  . &  . &  . &  . \\  
     . &  . &  . &  . &  . &  . &  . & 2a &  . &  . &  . &  . &  . &  . &  . &  . \\  
     . &  . &  . &  . &  . &  . &  . &  . & 2a &  . &  . &  . &  . &  . &  . &  . \\  
     . &  . &  . &  . &  . & -b &  . &  . & -b & 2a & -b &  . &  . & -b &  . &  . \\  
     . &  . &  . &  . &  . &  . & -b &  . &  . & -b & 2a & -b &  . &  . & -b &  . \\  
     . &  . &  . &  . &  . &  . &  . &  . &  . &  . &  . & 2a &  . &  . &  . &  . \\  
     . &  . &  . &  . &  . &  . &  . &  . &  . &  . &  . &  . & 2a &  . &  . &  . \\  
     . &  . &  . &  . &  . &  . &  . &  . &  . &  . &  . &  . &  . & 2a &  . &  . \\  
     . &  . &  . &  . &  . &  . &  . &  . &  . &  . &  . &  . &  . &  . & 2a &  . \\  
     . &  . &  . &  . &  . &  . &  . &  . &  . &  . &  . &  . &  . &  . &  . & 2a \\  
  \end{pmatrix}
  \begin{pmatrix}
  \phi_{0}  \\ \phi_{1}  \\ \phi_{2}  \\ \phi_{3}  \\ \phi_{4}  \\ \phi_{5} \\
  \phi_{6}  \\ \phi_{7}  \\ \phi_{8}  \\ \phi_{9}  \\ \phi_{10} \\ \phi_{11} \\
  \phi_{12} \\ \phi_{13} \\ \phi_{14} \\ \phi_{15} \\
  \end{pmatrix}
  \label{eqn_lap2d_dirx2_bc}
\end{equation}

Dirichet boundary conditions of this type are relevant to fully wall bounded
flows such as lid driven cavities \cite{smith1975comparative}.
However, the construction if \Cref{eqn_2dlap_oper} clearly produces the
wrong 2-dimensional operator.

%
\subsection{Dirichlet-Neumann boundary conditions}
\label{subsec-dir-neu}

A more typical CFD test case is the flow between two parallel plates
with a Dirichlet inflow and a Neumann outflow condition.
If in \Cref{fig-2dlattice}, the flow is from left to right,
then $L_{yy}$ is as in \Cref{eqn-lxx-dir} and

\begin{equation}
  L_{xx} = 
    \begin{pmatrix*}[r]
     a &  . &  . &  . \\
    -b &  a & -b &  . \\
     . & -b &  a & -b \\
     . &  . & -a &  a \\
  \end{pmatrix*}
  \label{eqn-lxx-neu}
\end{equation}

The corresponding 2D Laplace operator is:

\setcounter{MaxMatrixCols}{16}
\begin{equation}
  L\ket{\phi} = \\
  \begin{pmatrix}
    2a &  . &  . &  . &  . &  . &  . &  . &  . &  . &  . &  . &  . &  . &  . &  . \\  
    -b & 2a & -b &  . &  . &  . &  . &  . &  . &  . &  . &  . &  . &  . &  . &  . \\  
     . & -b & 2a & -b &  . &  . &  . &  . &  . &  . &  . &  . &  . &  . &  . &  . \\  
     . &  . & -a & 2a &  . &  . &  . &  . &  . &  . &  . &  . &  . &  . &  . &  . \\  
    -b &  . &  . &  . & 2a &  . &  . &  . & -b &  . &  . &  . &  . &  . &  . &  . \\  
     . & -b &  . &  . & -b & 2a & -b &  . &  . & -b &  . &  . &  . &  . &  . &  . \\  
     . &  . & -b &  . &  . & -b & 2a & -b &  . &  . & -b &  . &  . &  . &  . &  . \\  
     . &  . &  . & -b &  . &  . & -a & 2a &  . &  . &  . & -b &  . &  . &  . &  . \\  
     . &  . &  . &  . & -b &  . &  . &  . & 2a &  . &  . &  . & -b &  . &  . &  . \\  
     . &  . &  . &  . &  . & -b &  . &  . & -b & 2a & -b &  . &  . & -b &  . &  . \\  
     . &  . &  . &  . &  . &  . & -b &  . &  . & -b & 2a & -b &  . &  . & -b &  . \\  
     . &  . &  . &  . &  . &  . &  . & -b &  . &  . & -a & 2a &  . &  . &  . & -b \\  
     . &  . &  . &  . &  . &  . &  . &  . &  . &  . &  . &  . & 2a &  . &  . &  . \\  
     . &  . &  . &  . &  . &  . &  . &  . &  . &  . &  . &  . & -b & 2a & -b &  . \\  
     . &  . &  . &  . &  . &  . &  . &  . &  . &  . &  . &  . &  . & -b & 2a & -b \\  
     . &  . &  . &  . &  . &  . &  . &  . &  . &  . &  . &  . &  . &  . & -a & 2a \\  
  \end{pmatrix}
  \begin{pmatrix}
  \phi_{0}  \\ \phi_{1}  \\ \phi_{2}  \\ \phi_{3}  \\ \phi_{4}  \\ \phi_{5} \\
  \phi_{6}  \\ \phi_{7}  \\ \phi_{8}  \\ \phi_{9}  \\ \phi_{10} \\ \phi_{11} \\
  \phi_{12} \\ \phi_{13} \\ \phi_{14} \\ \phi_{15} \\
  \end{pmatrix}
  \label{eqn_lap2d_dir_neu}
\end{equation}

Now, only the corner nodes 0 and 12 on the inflow boundary are correct.
The interior nodes 5, 6, 9 and 10 are also still correct.
Nodes 1, 2 and 3 on the lower wall and 12, 14 and 15 on the upper wall
can be corrected as before.
At the inflow nodes 4 and 8 and the outflow nodes 7 and 11, there is a question
of whether to include the shear terms, i.e. the gradients in the y-direction.
It is usual to ignore these at the inlet where it is desirable to ensure that 
the specified boundary values are achieved.
For wall-bounded flows, the shear terms are usually retained at the outlet
to ensure boundary layers on the walls do not see a discontinuity in the
forces acting on the flow. 
At junctions of Dirichlet and Neumann boundary conditions, it is usual for
the Dirchlet condition to take precedence.
With these considerations the correct Laplacian is:

\setcounter{MaxMatrixCols}{16}
\begin{equation}
  L\ket{\phi} = \\
  \begin{pmatrix}
    2a &  . &  . &  . &  . &  . &  . &  . &  . &  . &  . &  . &  . &  . &  . &  . \\  
     . & 2a &  . &  . &  . &  . &  . &  . &  . &  . &  . &  . &  . &  . &  . &  . \\  
     . &  . & 2a &  . &  . &  . &  . &  . &  . &  . &  . &  . &  . &  . &  . &  . \\  
     . &  . &  . & 2a &  . &  . &  . &  . &  . &  . &  . &  . &  . &  . &  . &  . \\  
     . &  . &  . &  . & 2a &  . &  . &  . &  . &  . &  . &  . &  . &  . &  . &  . \\  
     . & -b &  . &  . & -b & 2a & -b &  . &  . & -b &  . &  . &  . &  . &  . &  . \\  
     . &  . & -b &  . &  . & -b & 2a & -b &  . &  . & -b &  . &  . &  . &  . &  . \\  
     . &  . &  . & -b &  . &  . & -a & 2a &  . &  . &  . & -b &  . &  . &  . &  . \\  
     . &  . &  . &  . &  . &  . &  . &  . & 2a &  . &  . &  . &  . &  . &  . &  . \\  
     . &  . &  . &  . &  . & -b &  . &  . & -b & 2a & -b &  . &  . & -b &  . &  . \\  
     . &  . &  . &  . &  . &  . & -b &  . &  . & -b & 2a & -b &  . &  . & -b &  . \\  
     . &  . &  . &  . &  . &  . &  . & -b &  . &  . & -a & 2a &  . &  . &  . & -b \\  
     . &  . &  . &  . &  . &  . &  . &  . &  . &  . &  . &  . & 2a &  . &  . &  . \\  
     . &  . &  . &  . &  . &  . &  . &  . &  . &  . &  . &  . &  . & 2a &  . &  . \\  
     . &  . &  . &  . &  . &  . &  . &  . &  . &  . &  . &  . &  . &  . & 2a &  . \\  
     . &  . &  . &  . &  . &  . &  . &  . &  . &  . &  . &  . &  . &  . &  . & 2a \\  
  \end{pmatrix}
  \begin{pmatrix}
  \phi_{0}  \\ \phi_{1}  \\ \phi_{2}  \\ \phi_{3}  \\ \phi_{4}  \\ \phi_{5} \\
  \phi_{6}  \\ \phi_{7}  \\ \phi_{8}  \\ \phi_{9}  \\ \phi_{10} \\ \phi_{11} \\
  \phi_{12} \\ \phi_{13} \\ \phi_{14} \\ \phi_{15} \\
  \end{pmatrix}
  \label{eqn_lap2d_dir_neu_bc}
\end{equation}

%
\subsection{Toplitz boundary conditions}
\label{subsec-toe}
The Toeplitz style of Laplace operator, \Cref{eqn-lxx-top}, does not
correspond to a commonly used CFD boundary condition and is not
analysed further.

\begin{equation}
  L_{xx} = 
    \begin{pmatrix*}[r]
     a & -b &  . &  . \\
    -b &  a & -b &  . \\
     . & -b &  a & -b \\
     . &  . & -b &  a \\
  \end{pmatrix*}
  \label{eqn-lxx-top}
\end{equation}

%
\subsection{Summary}
\label{subsec-2dlap-summary}

\begin{figure}[h]
    \centering
    \captionsetup{justification=centering}
    \includegraphics[clip, trim=2.0cm 8.0cm 2.cm 8.0cm,width=0.60\textwidth]{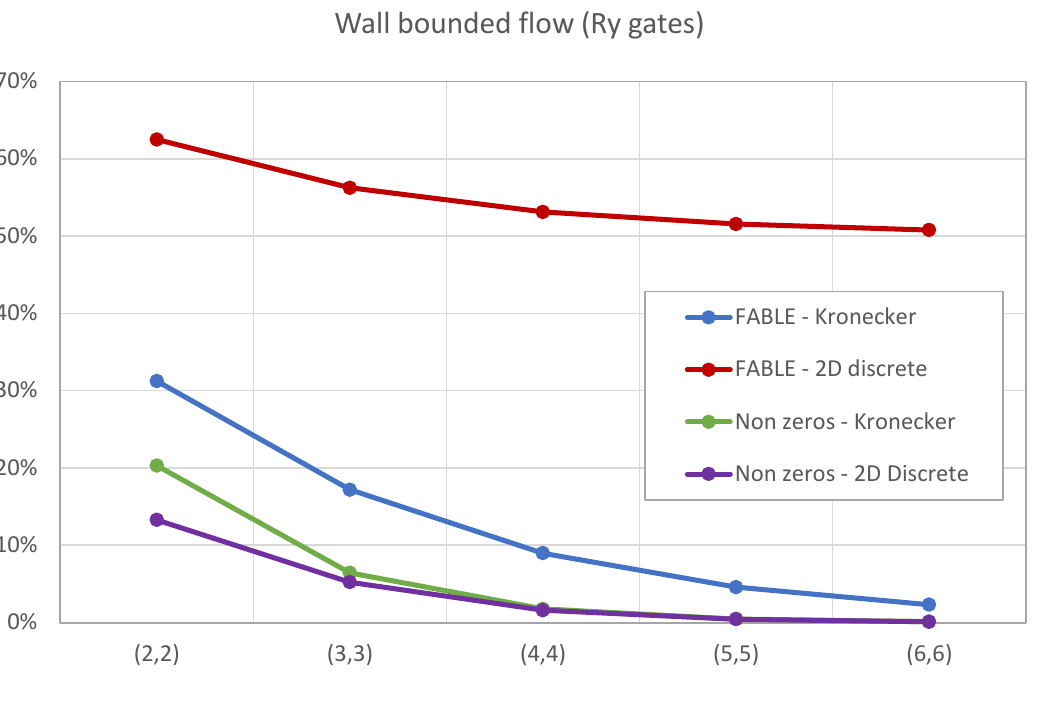}
    \caption{Percentage of rotation gates in a FABLE circuit for
    Dirichlet-Neumann boundary conditions using Kronecker sum and direct
    discretisation. (nx, ny) the are number of qubits in x and y directions.
    Also shown is percentage of non-zeros in each matrix formulation.}
    \label{fig_ksum_disc_appendix}
\end{figure}

\Cref{fig_ksum_disc_appendix} shows the impact of the matrix formulation on
the number or rotation gates in a FABLE circuit. The notation follows 
\cite{camps2022fable} where the abscissa is labelled (nx, ny) and
nx and ny are the number of qubits in x and y directions.
E.g., (6,6) corresponds to a 64x64 mesh and, hence, a 4,096x4,096
sparse matrix. 
The percentages are relative the to full circuit encoding of all
the entries in the matrix, i.e. 16,777,216 rotation gates for the
(6,6) case.
\Cref{fig_ksum_disc_appendix} also shows the percentage of non-zero
entries in the matrix. The effect of the different boundary treatments
is visible in the (2,2) case as the number of boundary nodes
is larger than the number of interior nodes. Otherwise the percentages
are dominated by the number of interiors nodes.

%% file: include/appendix-cluster.tex
%
\section{Clustering the mesh points}
\label{sec-clustering}

Mesh clustering is generally applied normal to a wall and can, therefore,
be considered as a 1-dimensional operation. Users generally wish to control
the near wall spacing, the rate of expansion away from the wall and the
spacing in freestream region where the mesh is uniform.
Additionally, the transition between the clustered and freestream regions
should be smooth. Depending on how the latter is defined can result in a
over-prescribed set of equations. 
L-QLES enforces smoothness at the expense of setting the near-wall 
spacing. This simplifies the solution of the clustering equations which
is preferred for the QLES experiments for which the code is intended.

The input parameters required for the mesh clustering are:
\begin{itemize}
    \item $L$, the total extent of the mesh,
    \item $n_t$, the total number of mesh points,
    \item $n_c$, the number of points in each clustered region,
    \item $r$, the expansion ratio in the cluster region,
    \item $C_{type}$, flag for 1 or 2 sided clustering.
\end{itemize}

For 2-sided clustering, the number of points in the uniform freestream
region, $n_u$ is:

\begin{equation}
 n_u = n_t - 2n_c + 2
\end{equation}

The $+2$ on the right side arise because the transition nodes between the
clustered and uniform regions are counted twice when solving for the mesh
distribution.

Defining $D$ as the freestream spacing, 
$C$ as the width of each clustered region and $d$ as the near wall spacing
gives:

\begin{equation}
  \sum_{i=0}^{n_c-2} r^i d  =  \frac{r^{n_c-1} - 1}{r-1} d =  C 
  \label{tests_clust01}
\end{equation}

\begin{equation}
  L = (n_u-1)D + 2C
  \label{tests_clust02}
\end{equation}

\begin{equation}
  r^{n_c-1} d =  D
  \label{tests_clust03}
\end{equation}

Where \Cref{tests_clust03} is the requirement that the transition from 
the clustered region to the uniform region is smooth. Solving the above
equations results in the mesh distribution:

\begin{equation}
  x =
  \begin{pmatrix}
    0  \\
    d \\
    d + rd \\
    \vdots \\
    C \\
    C + D \\
    \vdots \\
    C + (n_u-1)D \\
    \vdots \\
    L - rd \\
    L - d\\
    L
  \end{pmatrix}
\label{tests_clust05}
\end{equation}

\begin{figure}[h]
  \centering
  \captionsetup{justification=centering}
    \begin{subfigure}{0.45\textwidth}
    \includegraphics[width=\textwidth]{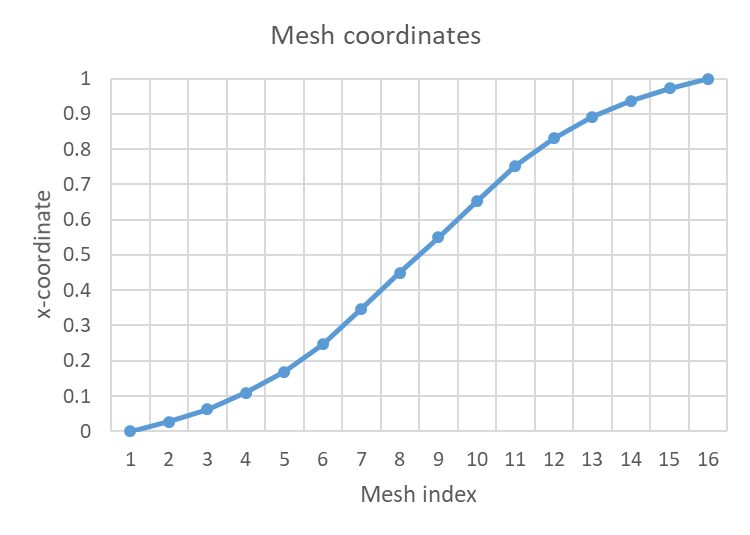}
    \caption{Mesh coordinates.}
    \label{fig-Poisson05a}
  \end{subfigure}
  \captionsetup{justification=centering}
    \begin{subfigure}{0.45\textwidth}
    \includegraphics[width=\textwidth]{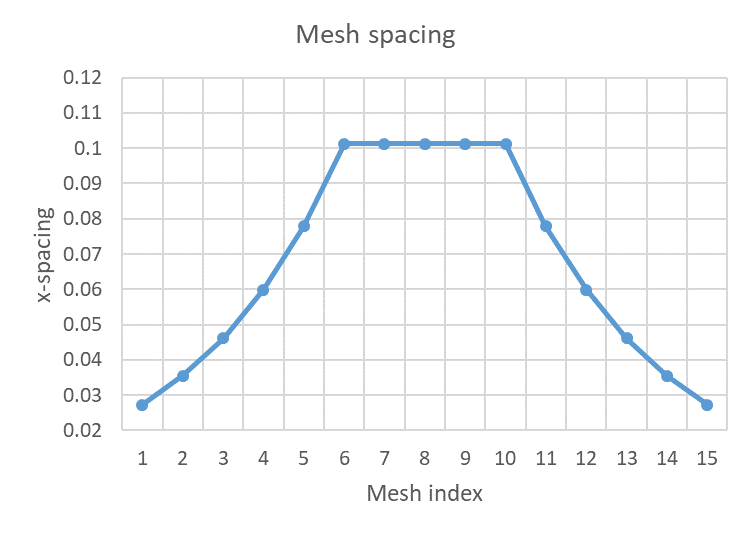}
    \caption{Mesh spacing.}
    \label{fig-Poisson05b}
  \end{subfigure}
  \caption{Mesh coordinates and spacing for $n_t = 16$, $n_c = 6$,
           $r=1.3$ and $L=1.0$.}
  \label{fig-cluster01}
\end{figure}

For 1-sided clustering, $n_u = n_t - n_c + 1$ and $L = (n_u-1)D + C$.
\Cref{fig-cluster01} shows a sample 2-sided mesh distribution and spacing.

%% file: include/appendix-lqles.tex
%
\section{Running L-QLES}
\label{sec-run-lqles}

The options for running L-QLES can be obtained by using the {\it -h} option on the 
command line to give:

\begin{verbatim}
l-qles.py -i <input file> {-c <x,y,z>} {-d} {-e} {-h} {-j} {-m} {-r} {-s}

     -i {name of input file}
     -c {x,y,z} cut slice of 3D solution to be plotted, default = x
     -d allow degnerate matrices, default = False
     -e calculate eigenvalues and condition number, default = False
     -h help menu
     -j split plots into separate windows for saving, default is single window
     -m plot matrix, default = False
     -r reorder matrix and RHS to use shell ordering of mesh, default = False
     -s plot solutons and mesh, default = False
\end{verbatim}

The structure of the input XML file is described in \Cref{sec-l-qles}.
The sample data files in the GitHub repository have additional comments
to aid understanding.

%
\subsection{Command line options}
\label{subsec-lqles-cmd}
The default for all options is \textit{off} unless otherwise stated.

\begin{description}
    \item[-i] Followed by the name of the XML input file. There is no default.
    \item[-c] Followed by {\it x}, {\it y} or {\it z}. For 3D cases this indicates which cutting
     plane is used to show the solution if the option \textbf{-s} is turned on.
     The cutting plane is position at the mid-point of the domain and default is an \textit{x}
     plane.
     
     \item[-d] As discussed in \Cref{subsec-testcases-2d-RRRR}, Laplacians that consist entirely of
     repeating and/or Neumann boundaries are degenerate. The default is to remove the degeneracy
     by applying a Dirichlet condition at a single point in the mesh determined by the input
     variable \textit{degfix}.
     Since other authors have used the degenerate form, this option does not apply the 
     degeneracy fix so that like with like comparisons can be made. Note that L-QLES cannot
     solve a degenerate Poisson equation and no solution file is stored in these cases.
     Other files have \textit{\_d} appended to their case name.
     
     \item[-e] Calculate the eigenvalues and condition number of the Laplacian. This scales
     with $O(N^3)$ where $N$ is the dimension of the matrix and so can only be used with
     small matrices. As \Cref{tab-illust1d} shows, L-QLES allows a wide range of condition
     numbers to be produced using small meshes.

     \item[-h] Display help menu.

     \item[-j] The default for the \textbf{-m} and \textbf{-s} plotting options is to
     create figures with 2 plots per pane. This option plots each figure separately which
     may be useful when preparing reports.

     \item[-m] Plot the matrix values and sparsity pattern using the Matplotlib 
     \cite{hunter2007matplotlib} \textit{imshow} and \textit{spy} options.

     \item[-r] Applies the shell reordering described in \Cref{subsec-reorder}. If this
     option is on, all files have \textit{\_r} appended to their case name.
     If the dengeneracy and reordering options are both on, then \textit{\_d\_r} is appended to
     the case name.

     \item[-s] Plot the mesh and contours of the solution variable. 
\end{description}

%
\subsection{Output files}
\label{subsec-lqles-ofiles}

Note the Laplacian, $L$, is normalised to have $||L||_{max}=1$ with the
same scale factor applied to the RHS state to ensure that the
solution state corresponds to the original problem. This does not
mean that the RHS state is normalised.

L-QLES outputs 2 types of files: Python and C/C++ compatible binary files:

\begin{description}
    \item[Laplacian] This is stored using the compressed sparse row format. 
    The name of the Python file is \textit{casename\_mat.npz} and the
    C/C++ binary file is \textit{casename\_mat.bin}.

    \item[RHS vector] The right-hand side vector contains the boundary values and
    the bulk inhomogeneous force term.
    The name of the Python file is \textit{casename\_rhs.npy} and the
    C/C++ binary file is \textit{casename\_rhs.bin}.

    \item[Solution vector] If the Laplacian is not degenerate, the solution
    vector is output.
    The name of the Python file is \textit{casename\_sol.npy} and the
    C/C++ binary file is \textit{casename\_sol.bin}.

    \item[Reordering matrix]  If the reordering option has been used then
    the column-wise permutation matrix ($Q$ in \Cref{eqn-reorder}) is written
    to the files \textit{casename\_ord.npz} and \textit{casename\_ord.bin}.
    As described in \Cref{subsec-reorder}, this matrix is needed to recover the
    solution to the original Laplacian from the solution to the reordered one.
\end{description}

If the\textbf{-r} and/or \textbf{-d} options have been used, the case name is
amended as described above.

%% file: main.bbl
\begin{thebibliography}{10}

\bibitem{hoefler2023disentangling}
T.~Hoefler, T.~H{\"a}ner, and M.~Troyer, ``Disentangling hype from practicality: on realistically achieving quantum advantage,'' {\em Communications of the ACM}, vol.~66, no.~5, pp.~82--87, 2023.

\bibitem{denton1997lessons}
J.~D. Denton, ``Lessons from rotor 37,'' {\em Journal of Thermal Science}, vol.~6, pp.~1--13, 1997.

\bibitem{reid1978design}
L.~Reid and R.~D. Moore, ``Design and overall performance of four highly loaded, high speed inlet stages for an advanced high-pressure-ratio core compressor,'' Tech. Rep. NASA-TP-1337, NASA, 1978.

\bibitem{levy2002summary}
D.~Levy, R.~Wahls, T.~Zickuhr, J.~Vassberg, S.~Agrawal, S.~Pirzadeh, and M.~Hemsch, ``Summary of data from the first aiaa cfd drag prediction workshop,'' in {\em 40th AIAA Aerospace Sciences Meeting \& Exhibit}, p.~841, 2002.

\bibitem{tinoco2018summary}
E.~N. Tinoco, O.~P. Brodersen, S.~Keye, K.~R. Laflin, E.~Feltrop, J.~C. Vassberg, M.~Mani, B.~Rider, R.~A. Wahls, J.~H. Morrison, {\em et~al.}, ``Summary data from the sixth aiaa cfd drag prediction workshop: Crm cases,'' {\em Journal of Aircraft}, vol.~55, no.~4, pp.~1352--1379, 2018.

\bibitem{lapworth2022hybrid}
L.~Lapworth, ``A hybrid quantum-classical cfd methodology with benchmark hhl solutions,'' {\em arXiv preprint arXiv:2206.00419}, 2022.

\bibitem{childs2017quantum}
A.~M. Childs, R.~Kothari, and R.~D. Somma, ``Quantum algorithm for systems of linear equations with exponentially improved dependence on precision,'' {\em SIAM Journal on Computing}, vol.~46, no.~6, pp.~1920--1950, 2017.

\bibitem{camps2022fable}
D.~Camps and R.~Van~Beeumen, ``Fable: Fast approximate quantum circuits for block-encodings,'' in {\em 2022 IEEE International Conference on Quantum Computing and Engineering (QCE)}, pp.~104--113, IEEE, 2022.

\bibitem{liu2021variational}
H.-L. Liu, Y.-S. Wu, L.-C. Wan, S.-J. Pan, S.-J. Qin, F.~Gao, and Q.-Y. Wen, ``Variational quantum algorithm for the poisson equation,'' {\em Physical Review A}, vol.~104, no.~2, p.~022418, 2021.

\bibitem{sato2021variational}
Y.~Sato, R.~Kondo, S.~Koide, H.~Takamatsu, and N.~Imoto, ``Variational quantum algorithm based on the minimum potential energy for solving the poisson equation,'' {\em Physical Review A}, vol.~104, no.~5, p.~052409, 2021.

\bibitem{ali2023performance}
M.~Ali and M.~Kabel, ``Performance study of variational quantum algorithms for solving the poisson equation on a quantum computer,'' {\em Physical Review Applied}, vol.~20, no.~1, p.~014054, 2023.

\bibitem{daribayev2023implementation}
B.~Daribayev, A.~Mukhanbet, and T.~Imankulov, ``Implementation of the hhl algorithm for solving the poisson equation on quantum simulators,'' {\em Applied Sciences}, vol.~13, no.~20, p.~11491, 2023.

\bibitem{sunderhauf2024block}
C.~S{\"u}nderhauf, E.~Campbell, and J.~Camps, ``Block-encoding structured matrices for data input in quantum computing,'' {\em Quantum}, vol.~8, p.~1226, 2024.

\bibitem{lapworth2022implicit}
L.~Lapworth, ``Implicit hybrid quantum-classical cfd calculations using the hhl algorithm,'' {\em arXiv preprint arXiv:2209.07964}, 2022.

\bibitem{harrow2009quantum}
A.~W. Harrow, A.~Hassidim, and S.~Lloyd, ``Quantum algorithm for linear systems of equations,'' {\em Physical review letters}, vol.~103, no.~15, p.~150502, 2009.

\bibitem{gilyen2018quantum}
A.~Gily{\'e}n, Y.~Su, G.~H. Low, and N.~Wiebe, ``Quantum singular value transformation and beyond: exponential improvements for quantum matrix arithmetics,'' {\em arXiv preprint arXiv:1806.01838}, 2018.

\bibitem{dong2021efficient}
Y.~Dong, X.~Meng, K.~B. Whaley, and L.~Lin, ``Efficient phase-factor evaluation in quantum signal processing,'' {\em Physical Review A}, vol.~103, no.~4, p.~042419, 2021.

\bibitem{darwish2016finite}
M.~Darwish and F.~Moukalled, {\em The finite volume method in computational fluid dynamics: an advanced introduction with OpenFOAM{\textregistered} and Matlab{\textregistered}}.
\newblock Springer, 2016.

\bibitem{cuthill1969reducing}
E.~Cuthill and J.~McKee, ``Reducing the bandwidth of sparse symmetric matrices,'' in {\em Proceedings of the 1969 24th national conference}, pp.~157--172, 1969.

\bibitem{aftosmis2004applications}
M.~Aftosmis, M.~Berger, and S.~Murman, ``Applications of space-filling-curves to cartesian methods for cfd,'' in {\em 42nd AIAA Aerospace Sciences Meeting and Exhibit}, p.~1232, 2004.

\bibitem{taylor1937mechanism}
G.~I. Taylor and A.~E. Green, ``Mechanism of the production of small eddies from large ones,'' {\em Proceedings of the Royal Society of London. Series A-Mathematical and Physical Sciences}, vol.~158, no.~895, pp.~499--521, 1937.

\bibitem{childs2012hamiltonian}
A.~M. Childs and N.~Wiebe, ``Hamiltonian simulation using linear combinations of unitary operations,'' {\em arXiv preprint arXiv:1202.5822}, 2012.

\bibitem{babbush2018encoding}
R.~Babbush, C.~Gidney, D.~W. Berry, N.~Wiebe, J.~McClean, A.~Paler, A.~Fowler, and H.~Neven, ``Encoding electronic spectra in quantum circuits with linear t complexity,'' {\em Physical Review X}, vol.~8, no.~4, p.~041015, 2018.

\bibitem{berry2015simulating}
D.~W. Berry, A.~M. Childs, R.~Cleve, R.~Kothari, and R.~D. Somma, ``Simulating hamiltonian dynamics with a truncated taylor series,'' {\em Physical review letters}, vol.~114, no.~9, p.~090502, 2015.

\bibitem{berry2018improved}
D.~W. Berry, M.~Kieferov{\'a}, A.~Scherer, Y.~R. Sanders, G.~H. Low, N.~Wiebe, C.~Gidney, and R.~Babbush, ``Improved techniques for preparing eigenstates of fermionic hamiltonians,'' {\em npj Quantum Information}, vol.~4, no.~1, pp.~1--7, 2018.

\bibitem{lin2022lecture}
L.~Lin, ``Lecture notes on quantum algorithms for scientific computation,'' {\em arXiv preprint arXiv:2201.08309}, 2022.

\bibitem{mottonen2004transformation}
M.~Mottonen, J.~J. Vartiainen, V.~Bergholm, and M.~M. Salomaa, ``Transformation of quantum states using uniformly controlled rotations,'' {\em arXiv preprint quant-ph/0407010}, 2004.

\bibitem{versteeg2007introduction}
H.~K. Versteeg and W.~Malalasekera, {\em An introduction to computational fluid dynamics: the finite volume method}.
\newblock Pearson education, 2007.

\bibitem{orszag1974numerical}
S.~A. Orszag, ``Numerical simulation of the taylor-green vortex,'' in {\em Computing Methods in Applied Sciences and Engineering Part 2: International Symposium, Versailles, December 17--21, 1973}, pp.~50--64, Springer, 1974.

\bibitem{comte1971simple}
G.~Comte-Bellot and S.~Corrsin, ``Simple eulerian time correlation of full-and narrow-band velocity signals in grid-generated,‘isotropic’turbulence,'' {\em Journal of fluid mechanics}, vol.~48, no.~2, pp.~273--337, 1971.

\bibitem{smith1975comparative}
R.~E. Smith~Jr and A.~Kidd, ``Comparative study of two numerical techniques for the solution of viscous flow in a driven cavity,'' {\em NASA Special Publication}, vol.~378, p.~61, 1975.

\bibitem{hunter2007matplotlib}
J.~D. Hunter, ``Matplotlib: A 2d graphics environment,'' {\em Computing in science \& engineering}, vol.~9, no.~03, pp.~90--95, 2007.

\end{thebibliography}
